\documentclass[twocolumn]{aastex61}
\pdfoutput=1 
\usepackage{amsmath,amstext}
\usepackage[T1]{fontenc}
\usepackage{apjfonts} 
\usepackage[figure,figure*]{hypcap}
\usepackage{amssymb}

\shorttitle{Evershed and counter Evershed flows in MHD simulations}
\shortauthors{A. L. Siu-Tapia et al.}

\begin{document}

\title{Evershed and counter-Evershed flows in sunspot MHD simulations}

\author{A. L. Siu-Tapia}
\affiliation{Max-Planck-Institut f\"ur Sonnensystemforschung, Justus-von-Liebig-Weg 3,
37077 G\"ottingen, Germany.}
\affiliation{Universit\"at G\"ottingen, Physikalisches Institut, Friedrich-Hund-Platz 1,
37077 G\"ottingen, Germany}

\author{M. Rempel}
\affiliation{High Altitude Observatory, NCAR, P. O. Box 3000, Boulder, CO 80307, USA.}

\author{A. Lagg}
\affiliation{Max-Planck-Institut f\"ur Sonnensystemforschung, Justus-von-Liebig-Weg 3,
37077 G\"ottingen, Germany.}

\author{S. K. Solanki}
\affiliation{Max-Planck-Institut f\"ur Sonnensystemforschung, Justus-von-Liebig-Weg 3,
37077 G\"ottingen, Germany.}
\affiliation{School of Space Research, Kyung Hee University, Yongin, 446-701 Gyeonggi, Republic of Korea.}

\begin{abstract}
There have been  a few reports in the literature of counter-Evershed flows observed
 in well developed sunspot penumbrae, i.e. flows directed towards the umbra along penumbral filaments. Here we investigate the driving forces of such counter-Evershed flows   in a radiative magnetohydrodynamic simulation of a sunspot and compare them with the forces acting on the normal Evershed flow. The simulation covers a timespan of 100 solar hours and generates an Evershed outflow exceeding $8$ km s$^{-1}$ in the penumbra along radially aligned filaments where the magnetic field is almost horizontal. Additionally, the simulation produces a fast counter-Evershed flow (i.e., an inflow near $\tau=1$) in some regions within the penumbra, 
reaching peak flow speeds of $\sim$12 km s$^{-1}$.
The counter-Evershed flows are transient and typically last a few hours before they turn into outflows again. 
By using the kinetic energy equation and evaluating its various terms in the simulation box, we found that the Evershed flow occurs due to overturning convection in a strongly inclined magnetic field while the counter-Evershed flows can be well described as siphon flows.
\end{abstract}

\section{Introduction}
The origin of large-scale flows in the penumbra of sunspots is of particular interest in observational and theoretical studies of sunspots. The most prominent flow in photospheric penumbrae is the Evershed flow \cite[EF;][]{Evershed1909}, an almost horizontal and radially outward directed flow of plasma with speeds in the kilometers per second range (typical spatially averaged speeds being $1-2$ km s$^{-1}$). The nearly horizontal flow is usually subsonic, although supersonic flows have been observed \citep[e.g.,][]{Wiehr1995, delToroIniesta2001, Bellot2004, Borrero2005}.
The physical mechanism responsible for driving the EF is closely connected to the fine structure of the penumbra, which is manifested through the penumbral intensity, magnetic field and velocity structure \cite[see, e.g., detailed reviews by][]{Solanki2003,Thomas2004,Thomas2008,Borrero2009,Scharmer2009,Schlichenmaier2009,Tritschler2009,Bellot2010,Borrero2011,Rempel2011a}.
All these quantities display an almost radial filamentary structure in the penumbra. In particular, the magnetic field configuration comprises two major components, one containing generally stronger and more vertical fields (so-called spines) which is thought to be the result of a protrusion of the umbral field into the penumbra \cite[see, e.g., review by][]{Borrero2011}, and the second one being composed by weaker and more inclined fields (intra-spines, hereafter referred to as filaments) where the EF takes place \cite[see, e.g.,][]{Tiwari2013}. 
This configuration has been referred to as \textit{uncombed penumbra} \citep{Solanki1993} or \textit{interlocking-comb structure} \citep{Thomas1992}.

Several models have been proposed to explain the filamentary nature of the penumbra, e.g., \citet{Danielson1961,Meyer1968,Choudhuri1986,Solanki1993,Schlichenmaier1998a, Schlichenmaier1998b,Thomas2002,Spruit2006,Scharmer2006}. However, not all the models contain a self-consistent description of the EF.
Some models based on stationary magnetic flux tubes representing the filaments, describe the EF as a siphon flow driven by a gas pressure difference between the footpoints of the flux tube \citep[e.g.,][]{Thomas1993}.
On the other hand, the dynamic magnetic flux tube model presented by \citet{Schlichenmaier1998a} produces an EF as a combination of hot plasma rising at the inner footpoint of the tube and a radial acceleration driven by a pressure gradient, consequence of radiative losses at the surface. There is also the model of \citet{Scharmer2006} that says that the EF takes place in field-free intrusions.
More recently, numerical magnetohydrodynamics (MHD) simulations have succeeded in reproducing the EF as a result of overturning convection in the presence of an inclined magnetic field \citep[e.g.,][]{Heinemann2007,Scharmer2008,Rempel2009a,Rempel2009b,Kitiashvili2009,Rempel2011b,Rempel2012}.
The EF is in these cases interpreted as the convective flow component in the direction of the magnetic field. In these models, the penumbral fine structure results from anisotropic magneto-convection.

During the early stages of penumbrae formation, line-of-sight (LOS) velocities of opposite sign with respect to that displayed by the typical EF have  been reported by \citet{Schlichenmaier2012} and \citet{Romano2014}. This has been interpreted as inflows towards the pore before the formation of the penumbra.
On rare occasions, well-developed penumbrae can also harbor counter-EF (inflows) at the photosphere \citep{Kleint2013,Louis2014,Siu2017}. 
In particular, \citet{Siu2017} reported the observation of  a prominent counter-EF with a lifetime of $\sim$2 days in the disk center-side of a well-developed penumbra. The counter-EF showed considerable fine structure, i.e., the counter-EF was confined along  "reversed"  penumbral filaments, with their heads/sources located at the outer penumbral boundary and their tails/sinks observed at the inner penumbral edge. \citet{Siu2017} showed that, as in the normal-EF, the filaments carrying the counter-EF display temperature and magnetic field gradients that are both consistent with the direction of the flow, being compatible with both the magneto-convective driver scenario as well as the siphon flow mechanism.

In this work, we analyze the results of a MHD high-resolution sunspot simulation  by \citet{Rempel2015}, which produces a penumbra with normal-EF (outflows) as well as a fast counter-EF (inflows) in some parts of the penumbra at photospheric heights, with lifetimes of several hours.
We investigate and identify the driving forces acting on both, the normal- and the counter-EF.

\section{Simulation}

\begin{figure*}[t]
    \centering
    \includegraphics[width=0.85\textwidth]{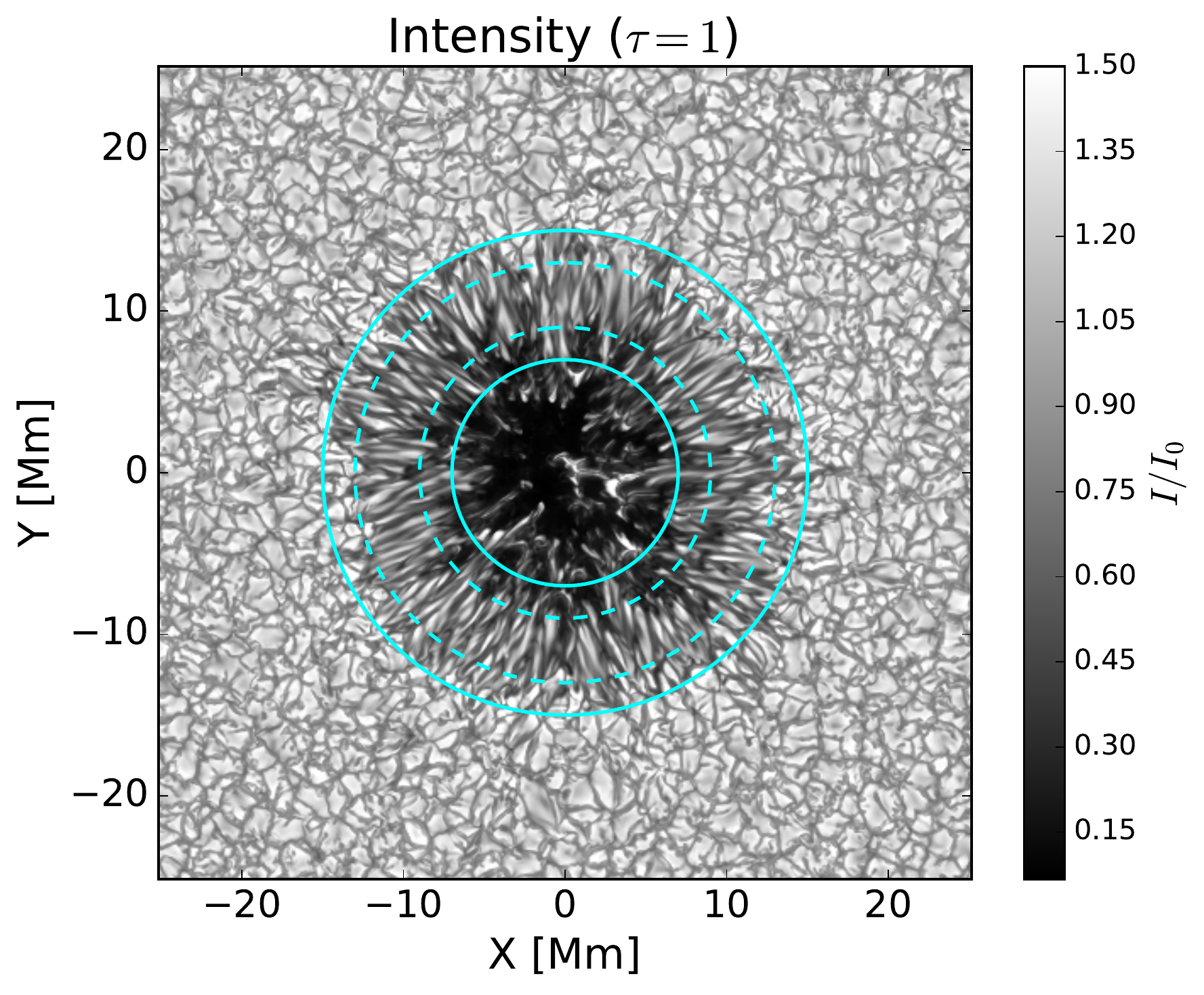} 
    \caption{Intensity image of the simulated sunspot, relative to the average quiet sun intensity $I_0$, at about 67 hours after initialization of the simulation run.  
The image corresponds to the full-spatial resolution series and shows the innermost $50.2\times 50.2$ Mm$^2$ of the simulation domain. Solid circles have been placed at radial distances of 7 Mm (where the azimuthally averaged intensity reaches $0.45 I_0$) and 15 Mm  (where the azimuthally averaged intensity reaches $0.9 I_0$) from the center of the sunspot, delimiting the inner and outer penumbra, respectively. The dashed circles at $R=9$ and 13 Mm delimit the middle penumbra.}\label{fig:1}
\end{figure*}

\begin{figure*}[t]
    \centering
    \includegraphics[width=\textwidth]{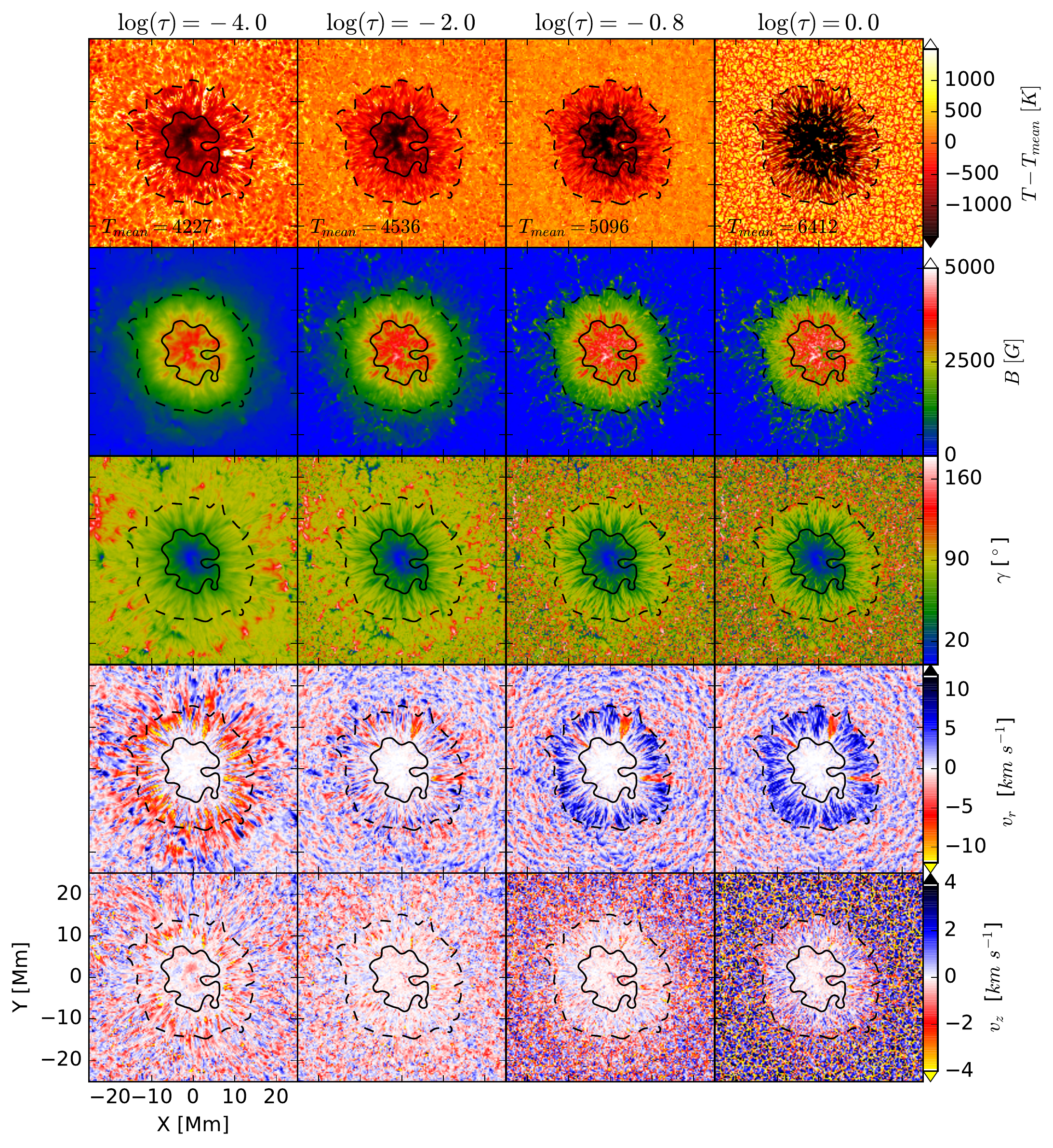} 
    \caption{Fine structure of the sunspot at different optical depth levels. From left to right: $\log(\tau)=-4, -2, -0.8$ and $0$. 
From top to bottom: Temperature perturbation, $T-T_{mean}$ [K]; magnetic field strength, $B$ [G]; field inclination, $\gamma$ [$^{\circ}$] with respect to vertical; radial flow velocity, $v_r$ [km s$^{-1}$]; and vertical flow velocity, $v_z$ [km s$^{-1}$]. 
A field inclination of $0^{\circ}$ corresponds to vertical field with the same polarity as the umbra, $90^{\circ}$ to horizontal, and $180^{\circ}$ to vertical field with opposite polarity of the umbra. Positive $v_r$ and $v_z$ values (blue-to-black colors) indicate outflows and upflows, respectively. Note that this sign convention differs from the one used in observational studies, which normally follow the spectroscopic definition of flow velocities along the line of sight, where negative values denote flows moving towards the observer and positive values flows moving away from the observer.
Black contour lines indicate the regions where the intensity $I<0.45 I_0$ and $I<0.9 I_0$. 
The images correspond to the same snapshot as Figure \ref{fig:1}.
}\label{fig:2}
\end{figure*}

We use the 3D high-resolution sunspot simulation setup described in detail by \citet{Rempel2015} and used therein to study the role of the penumbra and associated flows for sunspot decay. 
The simulation is based on the MURaM radiative MHD code \citep{Vogler2005,Rempel2009b,Rempel2014} and covers a timespan of 100 solar hours. An animation covering 25 hr (from $t=50$ to 75 hr after the initialization of the simulation) is provided as part of the online material of \citet{Rempel2015}.

As described in \citet{Rempel2015}, the sunspot setup used open boundary conditions that do not maintain the initial field structure against decay driven by convective motions. The simulation initial state was a relaxed small scale dynamo, the sunspot simulation was then initialized by inserting an axisymmetric self-similar magnetic field structure into the domain with total initial flux of $9\times10^{21}$ Mx, a field strength at the bottom of the domain of around 20 kG and at the top of the domain of around 3 kG.
The simulated sunspot had an almost constant magnetic flux  for the simulated timespan of 100 hours and used a magnetic top boundary condition that imposes a sufficiently horizontal field to maintain a penumbra.

After $t=50$ hours, the simulated sunspot showed well-developed penumbral fine structure (see, e.g., Figure \ref{fig:1} for an intensity image obtained about 67 solar hours
after initialization of the simulation), i.e., radially aligned filaments with close to horizontal field (see Figure \ref{fig:2}). Along these filaments, there are fast radial outflows reaching peak flow velocities of $\sim12$ km s$^{-1}$.
In addition to these  outflow regions, there are also some patches in the penumbra that have a counter-EF (inflows in the photosphere). These regions are transient, lasting a few hours before they turn into outflows again.

To investigate the nature of the counter-EF, we analyze the simulation time-steps from 60 to 70 solar hours (The range of time during which the counter-EF are found in the simulations).
As described in \citet{Rempel2015}, the photosphere is located about 700 km beneath the top boundary, and the simulation domain extends about 18 Mm in depth below the photosphere and 98 Mm horizontally, using a grid spacing of 24 km vertically and 48 km horizontally, which is required to capture penumbral fine structure and the Evershed flow.
The simulation ran on a $2048\times2048\times768$ $xyz$-grid (in the following presentation the $z$-direction is vertical). 
Full resolution data cubes were written every 4500 seconds, whereas data cubes at half the spatial resolution
 were written every 900 seconds. A non-gray run was restarted from one of the full resolution cubes 
and evolved for 5000 time steps, i.e. 1125 seconds. All of the snapshots have improved numerics to address the drift problem described in \citet{Rempel2015}, related to numerical diffusion and need for $\nabla \cdot B$ cleaning.

A full-spatial-resolution snapshot obtained with non-gray radiative transfer 
($t\sim67$ solar hours) is used in this work to study the filamentary structure of the  normal- and the counter-EF.
The  half-resolution data cubes obtained with gray radiative transfer
are employed to analyze the driving of the flows in the penumbra and their evolution in time.


\section{Results}
\subsection{Filamentary structure of the penumbra}

Figure \ref{fig:1} shows the normalized intensity of the simulated sunspot at $\tau=1$ for a snapshot at $t\sim67$ solar hours. 
Here we are mainly interested in the penumbra, which appears to have a uniform filamentary structure in intensity at $\tau=1$.

Figure \ref{fig:2} displays the filamentary fine structure of the penumbra as seen in the different physical quantities and at different optical depths: $\log(\tau)=-4,-2,-0.8$ and $0$. The filamentary structure of the penumbra is more evident at $\log(\tau)=0$, where the penumbral filaments appear as bright (hot) elongated channels with magnetic fields that become more inclined from the inner penumbral edge outwards, and return back to the interior ($\gamma>90^{\circ}$) towards the outer penumbral edge. Also, the velocity vector in the penumbral filaments is mostly radial.

The radial velocity maps in Figure \ref{fig:2} 
(fourth row) show that the penumbra is dominated by an inflow with respect to the center of the spot (red-to-yellow colors) at $\log(\tau)=-4.0$,
which resembles the chromospheric \textit{inverse} Evershed flow (IEF) \citep{Dialetis1985} but occurs at lower heights, and by an outflow below $\log(\tau)=-2.0$ (blue-to-black colors), which represents the photospheric normal-EF.
However, there are also some regions within the penumbra at $\log(\tau)=-2.0, -0.8$, and $0$ where the radial velocity is negative, indicating photospheric counter-EF.
 These photospheric inflows 
occur along penumbral filaments. Their associated vertical velocity shows large negative values (downflows) close to the inner penumbral boundary, at the end of these inflow filaments. 

\begin{figure*}[t]
    \centering
   
	\includegraphics[width=0.95\textwidth]{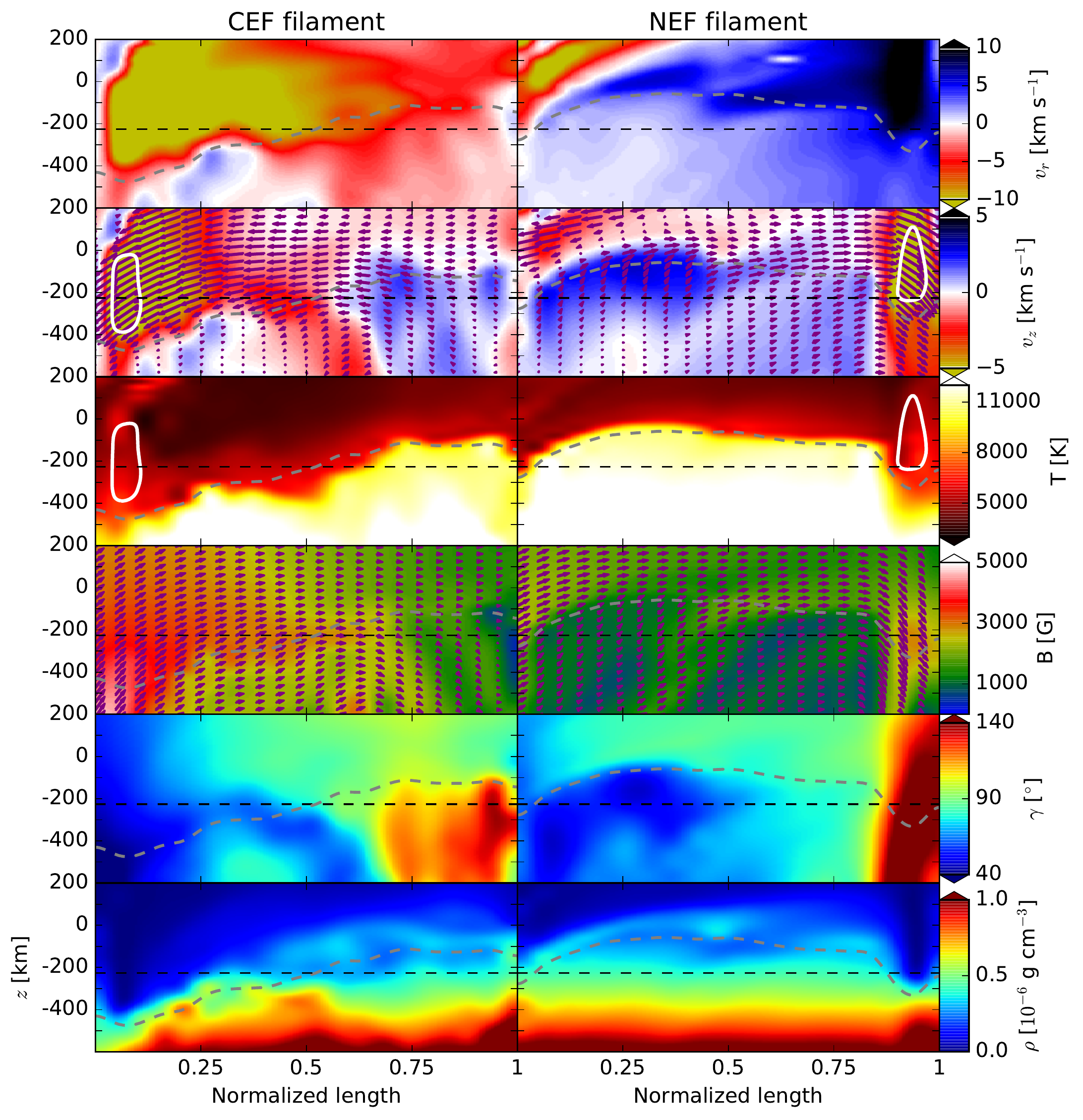} 
    \caption{Vertical cross section through the central axes of  two filaments, 
reaching from $-600$ km to 200 km above the surface: (left) CEF-carrying filament  and (right) NEF-carrying filament. 
 The filaments were selected from the snapshot at $t\sim67$ solar hours, which corresponds to a time when the CEF is prominent and stable. 
Panels show, from top to bottom: radial flow velocity $v_r$, vertical flow velocity $v_z$, temperature $T$, magnetic field intensity $B$, field inclination with respect to vertical $\gamma$, and density $\rho$. 
Gray dashed lines indicate the $\tau=1$ levels. Horizontal dashed lines  are placed at $z=-226$ km, which is the average height of $\tau=1$ in the penumbra with respect to the quiet Sun. The arrows in the $v_z$  and $B$ maps indicate the direction of the flow and magnetic field, respectively, mainly to guide the
eye.  The white contour lines on the $v_z$ and $T$ maps indicate sink regions with supersonic downflows, i.e. where $v_z<-8$ km s$^{-1}$.
As a reference, the sound speed is about 6 km s$^{-1}$ close to the $\tau=1$ surface. The orientation of the panels is such that the umbra  is located to the left.}\label{fig:4}
\end{figure*}

 In order to get insight into the physical differences between the regions harboring the counter-EF (CEF) and the regions with the normal-EF (NEF), in Figure \ref{fig:4} we compare the 
 characteristics of the gas and the magnetic field along the central axes of two radial flow channels which carry a CEF and a NEF, respectively. 
Figure \ref{fig:4} shows a vertical cross section along the normalized central  axes of the two selected penumbral filaments, a CEF-carrying filament (left panels) and a NEF-carrying filament (right panels), from $z=-600$ km  to 200 km above the surface (which is defined as the average height of the $\tau=1$ level in the quiet Sun). 
The selected individual filaments are representative of each filament type (CEF-carrying and NEF-carrying, respectively) since  most of the qualitative physical characteristics discussed below are true for most of the filaments in this snapshot.

The velocity panels in Figure \ref{fig:4} show the sources of the NEF, i.e. the portion of the filament on the right panels that harbors both upflows ($v_z>0$) and outflows ($v_r>0$), spanning around  $85\%$ of the filament's normalized length, while the sinks (regions where $v_z<0$ and $v_r>0$)  are observed  towards the outermost $15\%$ of the filament's length. 
In contrast, the sources of the CEF (regions where $v_z>0$ and $v_r<0$ on the left panels) are located in the outer endpoint of the filament (i.e., in the endpoint closer to the quiet Sun) from $\sim0.6$ to 1 in the normalized length scale, while their sinks (regions where $v_z<0$ and $v_r<0$) dominate mainly within the innermost $30\%$ of the filament. 
Therefore, unlike the CEF-carrying filament, almost  the entire NEF-carrying filament behaves as a source of flow.

Nonetheless, there are also some similarities. In both the CEF and NEF flow channels, the upflows  harbor hot gas   ($T>10000$ K) and are associated with relatively weak magnetic fields. The magnetic field in the upflow cells  becomes more inclined with height and is aligned with the direction of the flows (see arrows on $v_z$ and $B$ panels in Figure \ref{fig:4}). 
Furthermore, in both the NEF and CEF, the gas flows  are supersonic in the sink regions and exhibit lower temperatures than at their sources ($T<8000$ K). The downflowing gas at the sink regions is slightly warmer than in its surroundings, and the downflow speeds are generally supersonic at these locations, dropping rapidly with depth from supersonic to subsonic. These properties suggest that for both the NEF and CEF, the deceleration takes place in the form of shocks.

Compared to the field strength at the source regions ($B<2000$ G) in both, CEF and NEF, 
the magnetic field presents a strengthening 
at the sink regions ($B > 2500$ G), 
 where it gradually becomes more vertical in the flow direction.
Furthermore, in both NEF and CEF, the magnetic field polarity in the sink regions is opposite to that at their corresponding sources.
Because  in Figure \ref{fig:4} we concentrate on the central axes of the filaments, we see only the `end of filament' sinks. Nonetheless, most of the overturning mass flux goes to the lateral sinks, so the end of filament sinks represent only a part of the picture since a significant fraction ($\sim50\%$) of the returning mass flux is found in regions with still upward pointing field when the lateral sinks are included.

Dashed gray  lines in Figure  \ref{fig:4} indicate the variation in height of the $\tau=1$ levels along the two  flow channels with respect to the reference penumbral average height $z=-226$ km (horizontal black dashed lines) .
Such constant optical depth levels are depressed in the downflowing part of the filaments and are elevated in the upflowing part of the filaments. 
 The depression of the constant optical depth surfaces in the sink regions are caused by lower temperatures and densities of the downflowing gas compared to the upflowing gas. 
This is an important aspect to consider in observational studies of penumbral filaments, in which the dynamics of the flows and the magnetic field structure are only accessible at constant optical depth surfaces, see also discussion in \citet{vannoort2013}.
In this simulation, the magnetic field strength suffers a large enhancement in the downflowing part of the filaments (sinks) along the $\tau=1$ level, taking values of up to $\sim5$ kG in the  CEF filament, and $\sim2.5$ kG in the  NEF-carrying filament. 
For the CEF filaments, such a very large  strengthening of the field is partly the result of the strong depression of the $\tau=1$ level at the sinks and their close vicinity to the umbral field (note that the CEF sinks are mainly located close to the umbral boundary).
The strengthening of the field at the sinks of the CEF filaments at  constant geometrical height also contributes (as can be seen in the left column of Figure \ref{fig:4}).
Moreover, the sinks of the NEF filaments also present a field intensification near the $\tau=1$ level. 
Such local field strengthening 
might be produced by the supersonic downdrafts of magnetized flux concentrations at the sink regions as suggested by \citet{vannoort2013} for explaining the observation of field strengths of the order of 7 kG in supersonic downflow regions at the outer penumbra of a sunspot.

\subsection{Driving forces of the penumbral flows. }\label{sec:3.3}

Our following analysis follows very closely the analysis performed by \citet{Rempel2011b} aimed to investigate the physical processes that lead to the driving of large-scale outflows around sunspots.
Therefore, in order to investigate which forces are responsible for driving the inflows and the outflows in the penumbra, we analyze the various terms in 
the kinetic energy equation used by MURaM, which is derived from the momentum equation. 
Since we concentrate on time averages within  this section, we use the following energy balance equation, which neglects the temporal derivatives and assumes stationary flows:

\begin{equation}
\underbrace{\vec{v}\cdot (\rho \vec{g}-\nabla{p})}_\text{Pressure}+\underbrace{\vec{v}\cdot (\vec{j}\times\vec{B})}_\text{Lorentz}\underbrace{-\rho \vec{v}\cdot [(\vec{v}\cdot \nabla)\vec{v}]}_\text{Acceleration}+\underbrace{\vec{v}\cdot \vec{F_{visc}}}_\text{Viscosity}=0
\label{eq:1}
\end{equation}

In this energy balance equation, a negative acceleration term represents a source of kinetic energy given that, under the assumption of stationarity, the acceleration term is identical to the negative divergence of the kinetic energy flux, $\rho \vec{v} v^2/2$.

We compute the individual terms in the energy balance equation as follows: 

 \begin{equation}
\langle P_i^{\pm}\rangle=\langle v_i^{\pm}[\rho g_i-(\nabla p)_i] \rangle 
\label{eq:2}
\end{equation}
\begin{equation}
\langle L_i^{\pm}\rangle=\langle v_i^{\pm}(\vec{j}\times \vec{B})_i] \rangle
\label{eq:3}
\end{equation}
\begin{equation}
\langle A_i^{\pm}\rangle=\langle -\rho v_i^{\pm}[(\vec{v} \cdot \nabla)\vec{v}]_i \rangle
\label{eq:4}
\end{equation}
The viscosity term is not explicitly calculated but instead we use an approximated magnitude which we call \textit{residual force}:
\begin{equation}
\langle R_i^{\pm}\rangle=\langle -(P_i^{\pm}+L_i^{\pm}+A_i^{\pm})\rangle
\label{eq:5}
\end{equation}

\begin{figure*}[t]
    \centering
 \includegraphics[width=0.45\textwidth]{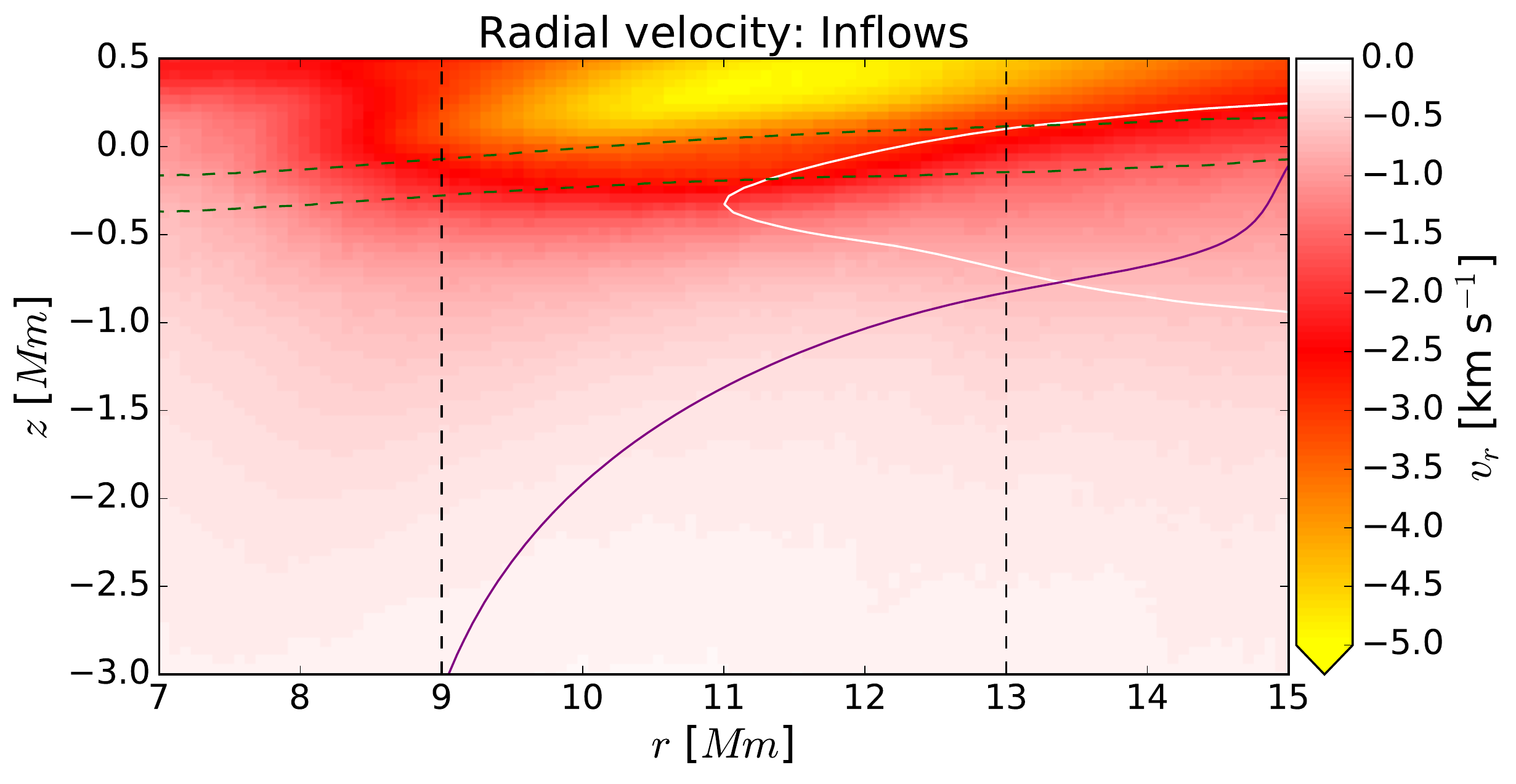} 
 \includegraphics[width=0.45\textwidth]{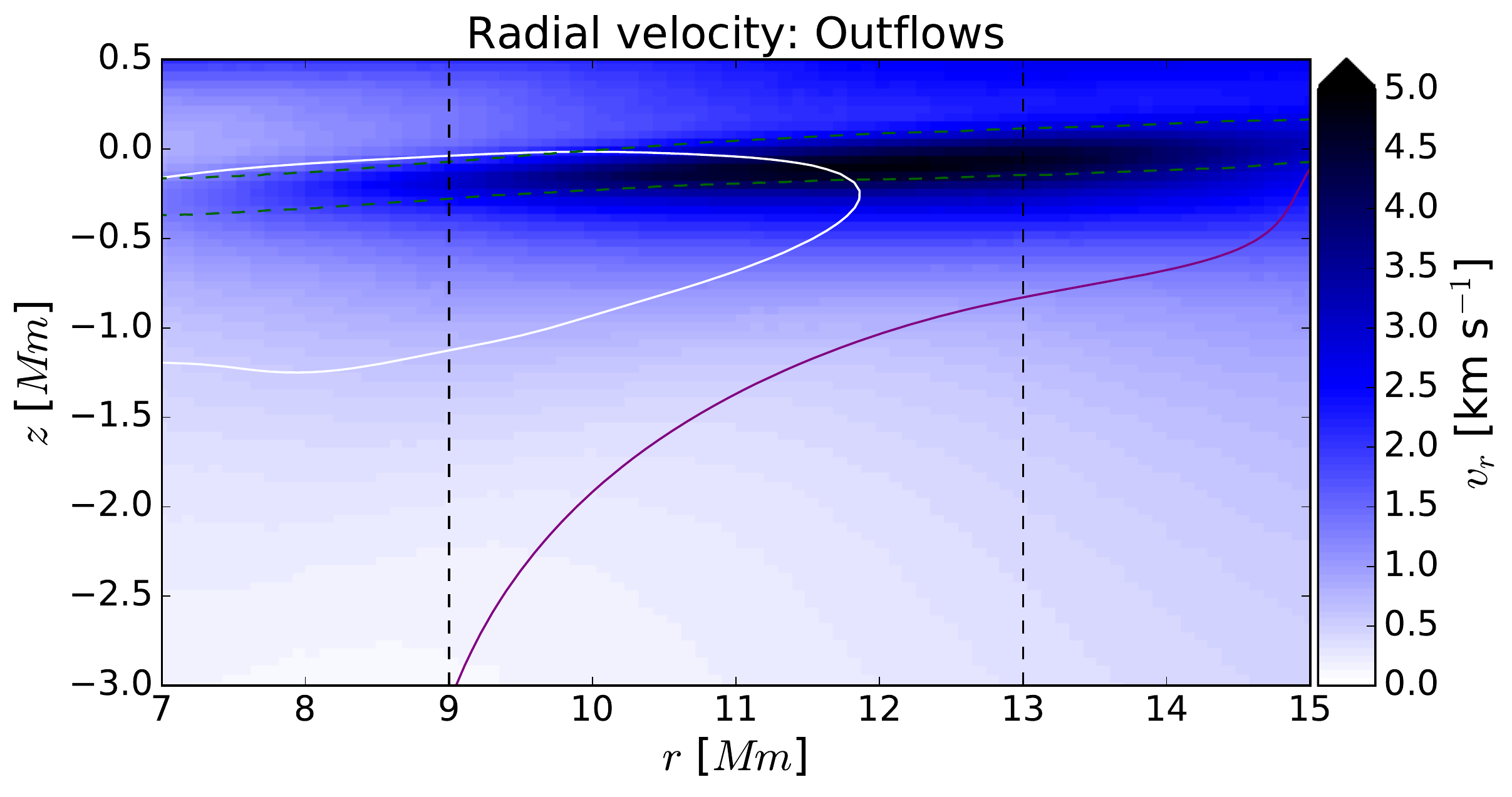} 
    \includegraphics[width=0.45\textwidth]{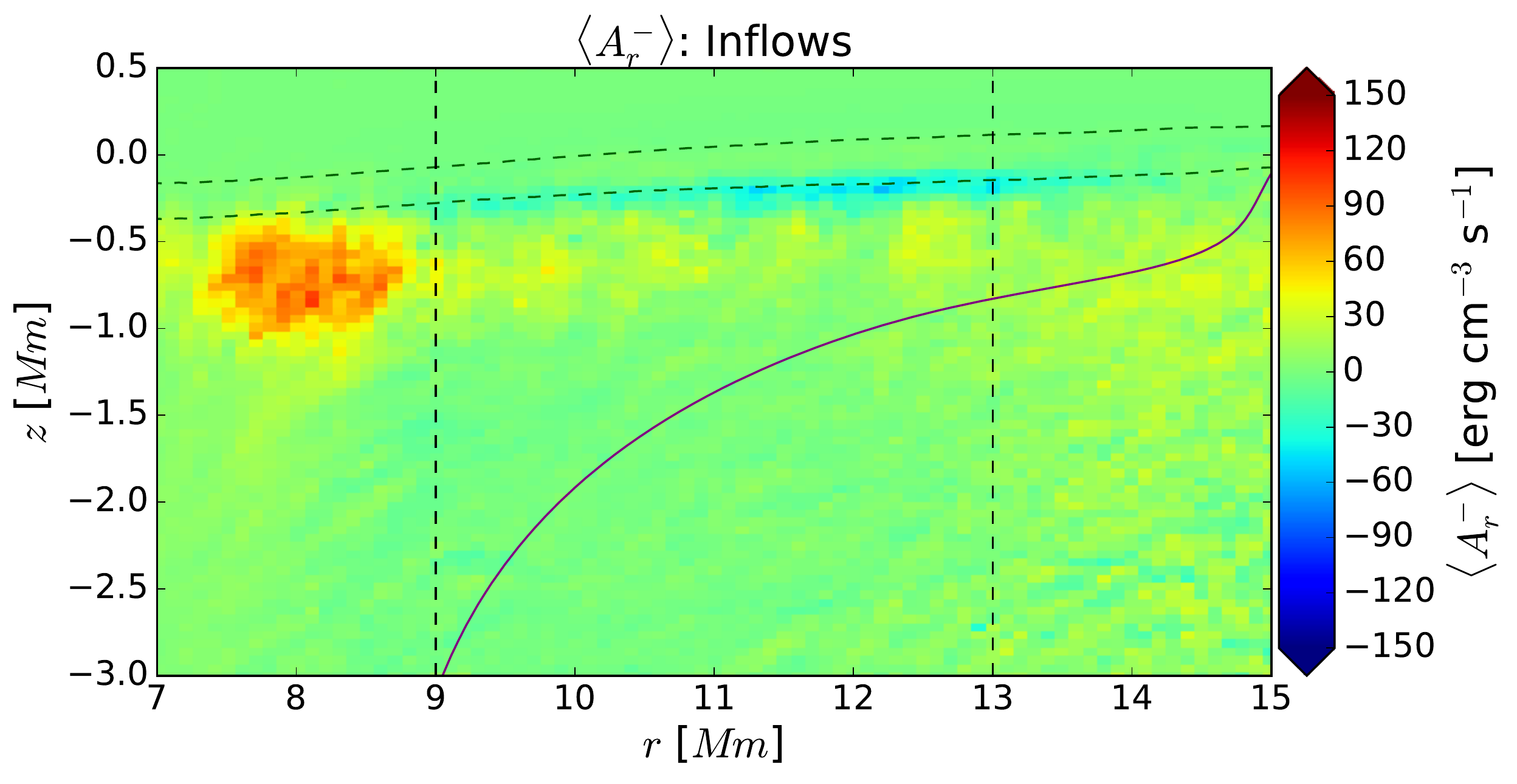} 
	  \includegraphics[width=0.45\textwidth]{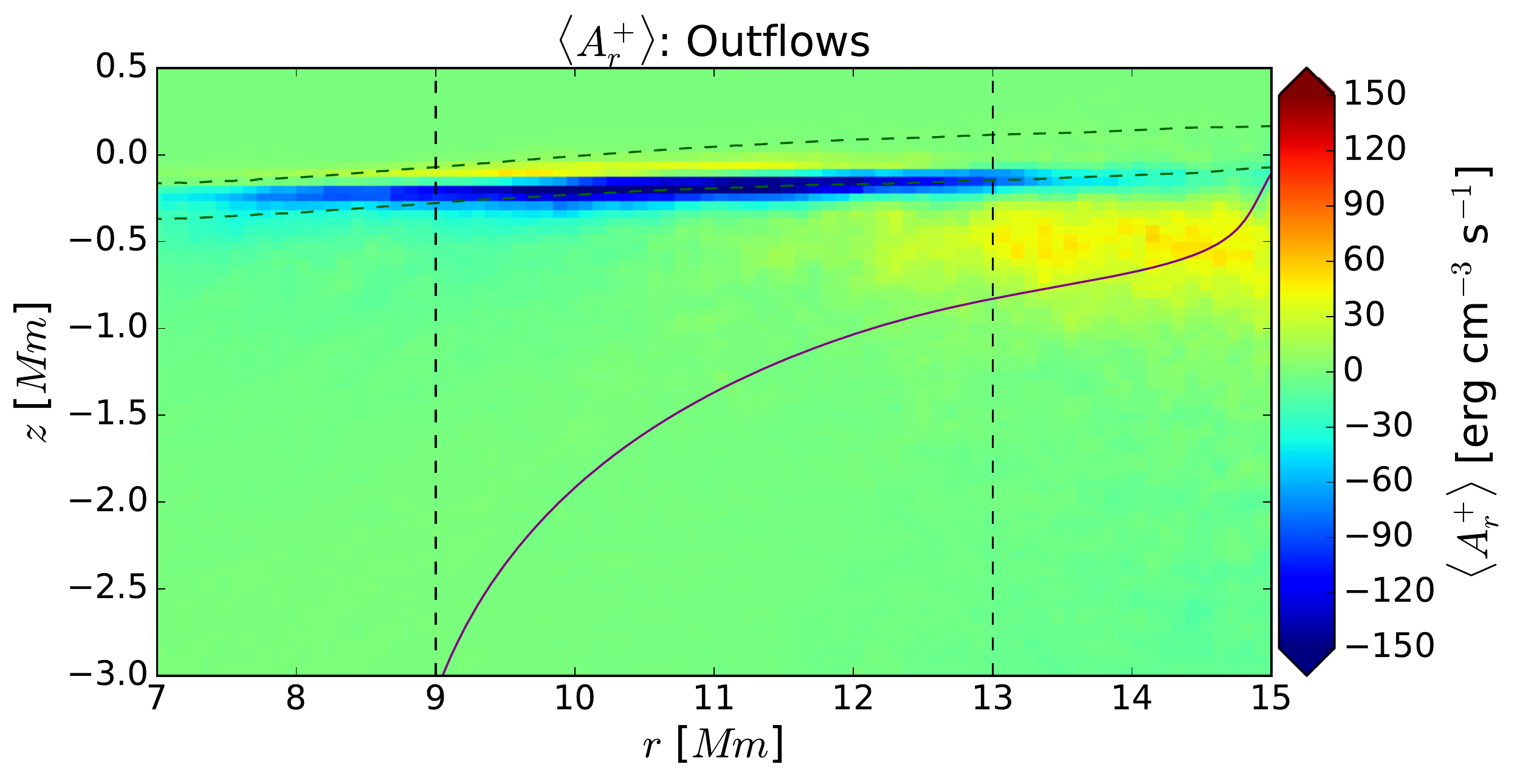} 
\includegraphics[width=0.45\textwidth]{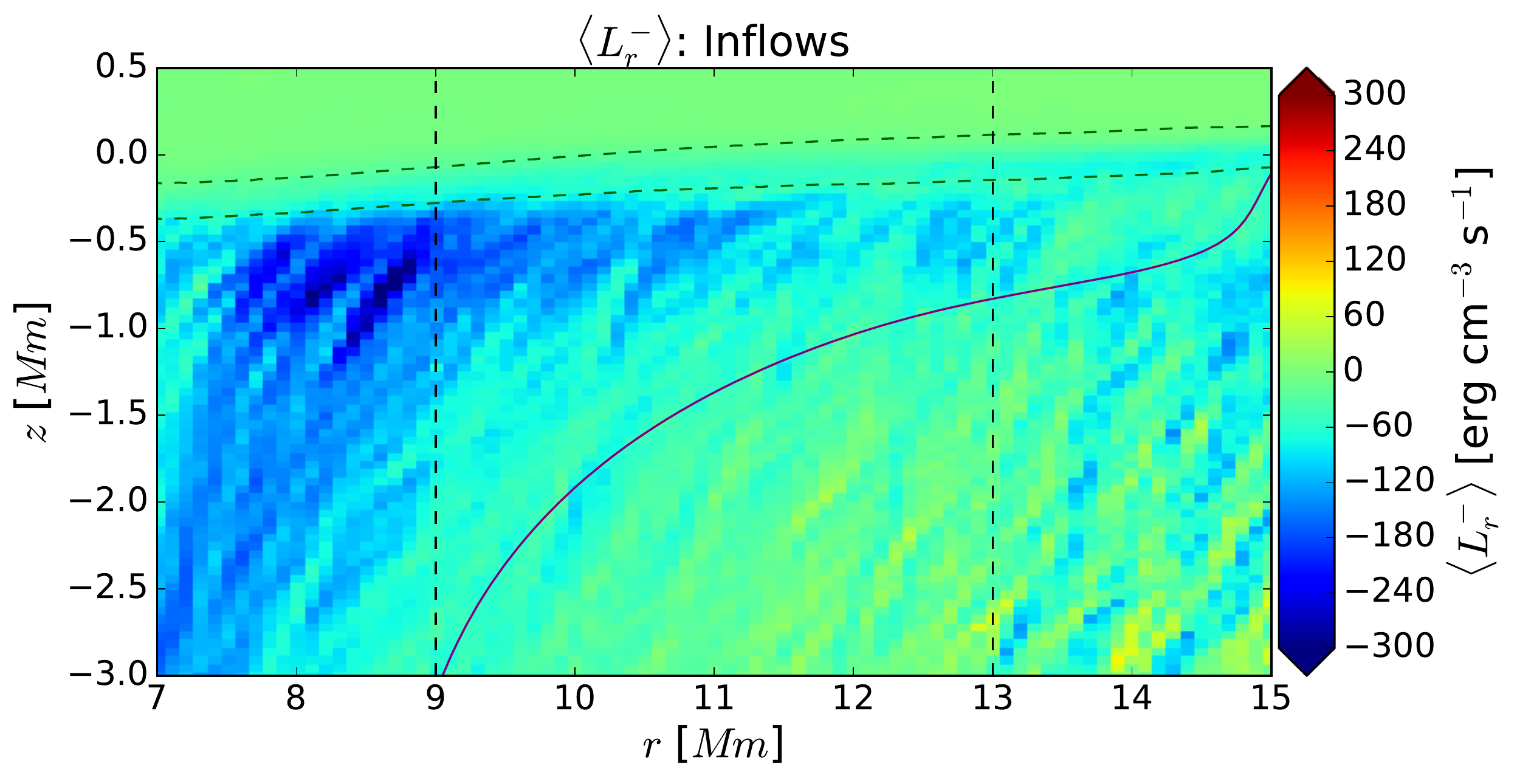} 
	  \includegraphics[width=0.45\textwidth]{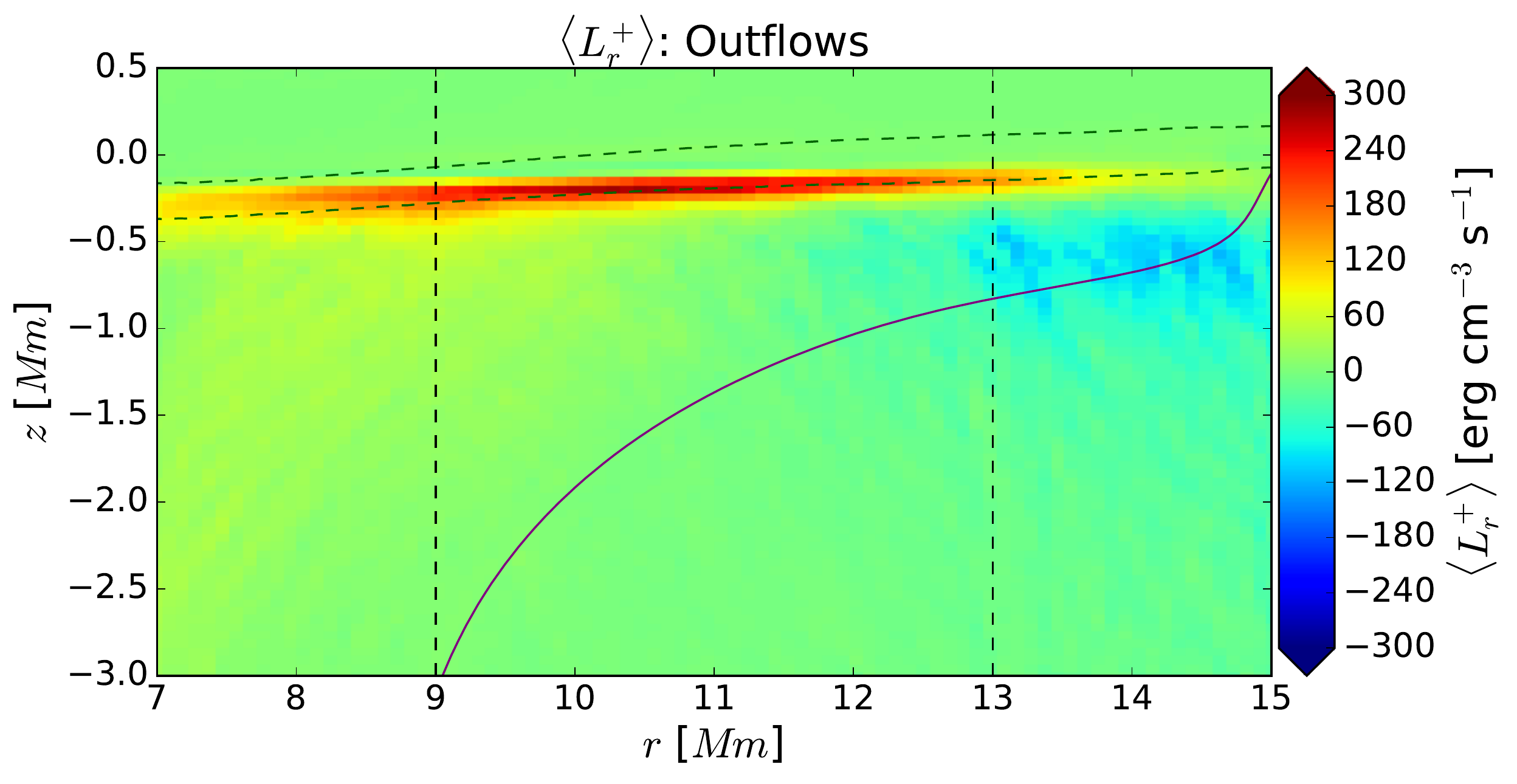} 
\includegraphics[width=0.45\textwidth]{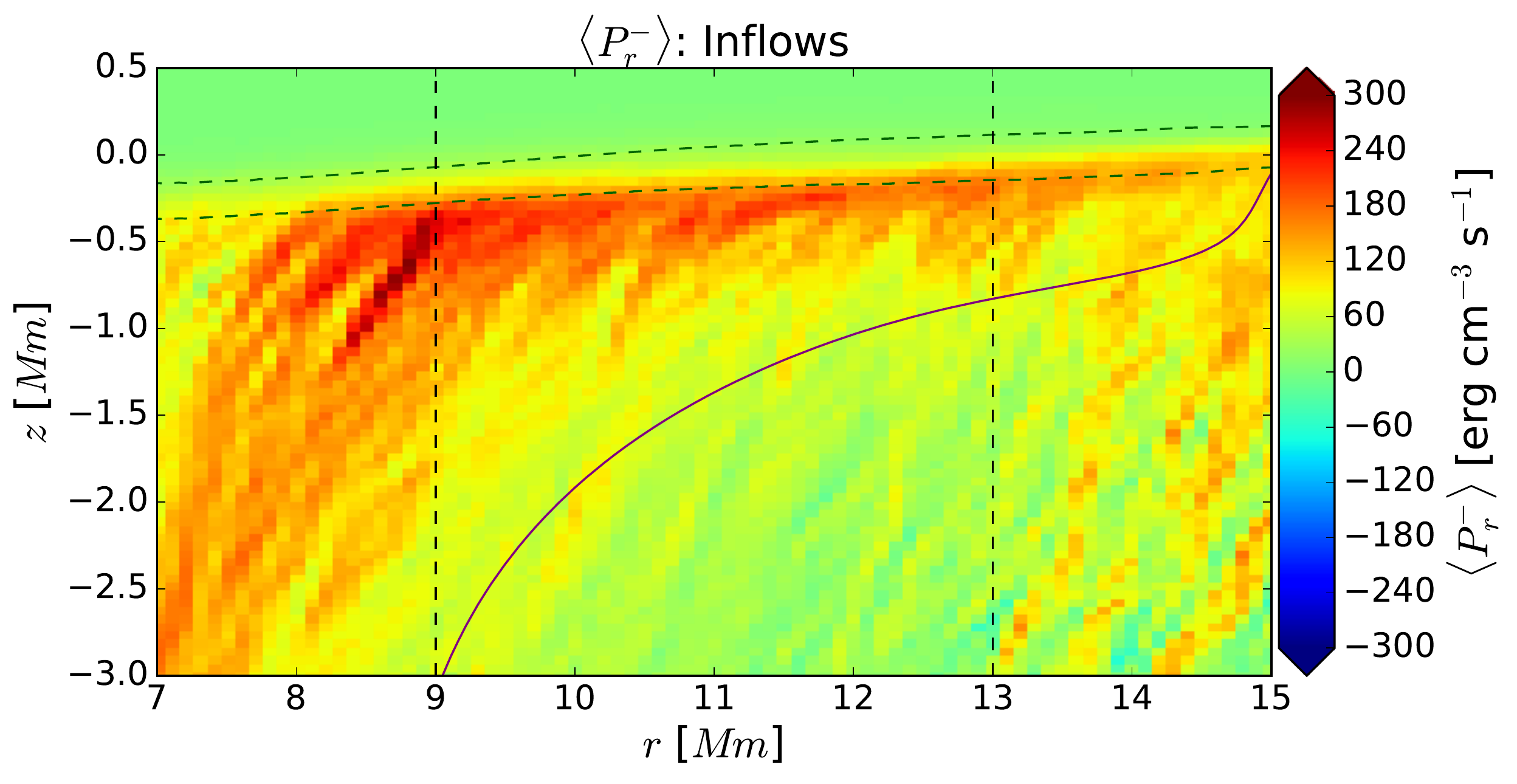} 
	  \includegraphics[width=0.45\textwidth]{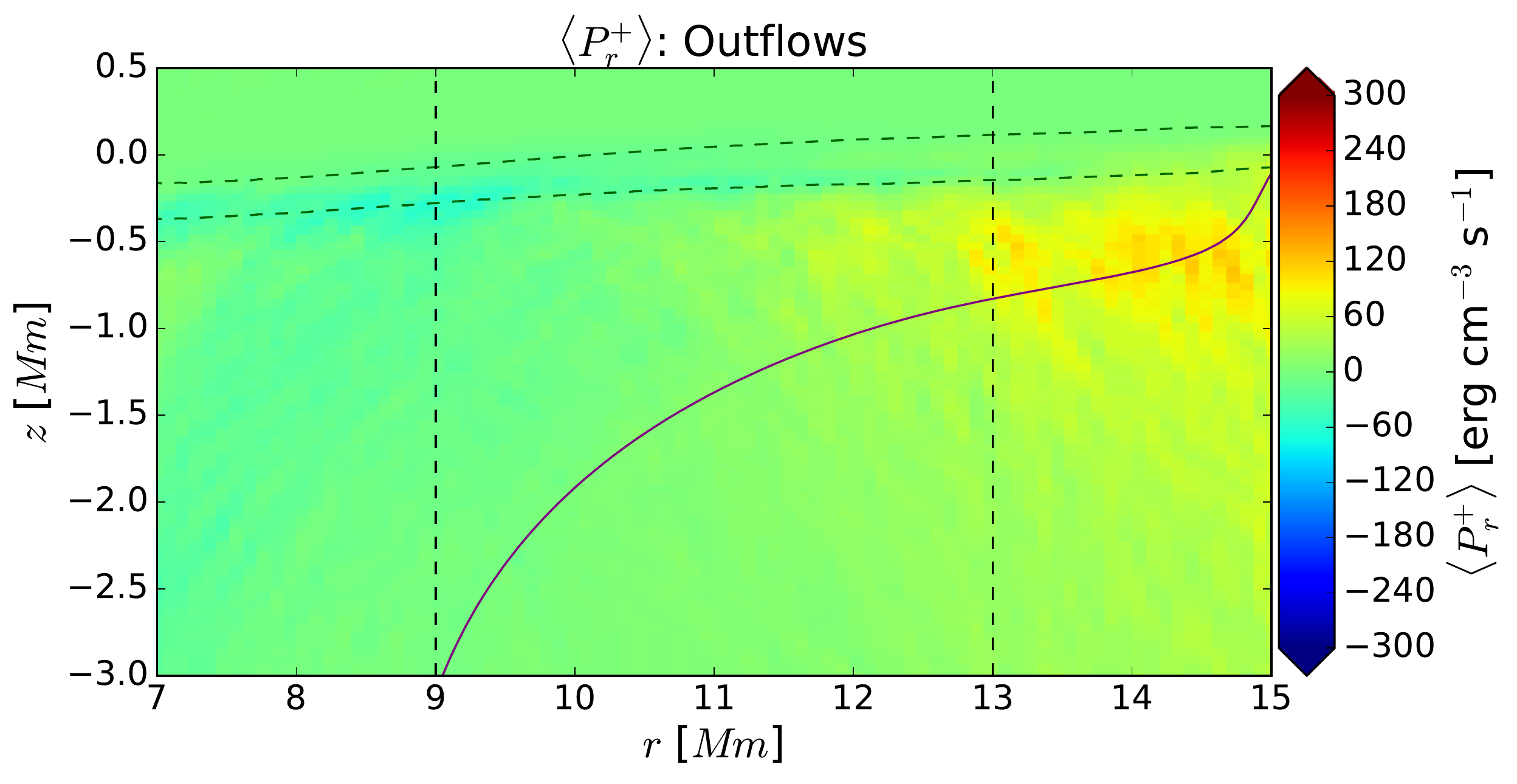} 
    \caption{Radial flow velocity and radial energy conversion terms as functions of radius and height, using masks that select only inflows ($v_r<0$, left plots) and only outflows ($v_r>0$, right plots)  in the penumbra. All plots were averaged from 60 to 70 solar hours using the half spatial resolution 3D cubes with temporal cadence of 900 sec.
From top to bottom: radial flow velocity, radial components of the acceleration, Lorentz and pressure forces. 
 The residual terms in both cases are negligible.
Vertical dashed lines delimit the inner penumbra ($r=7-9$ Mm from the center of the sunspot), the middle penumbra ($r=9-13$ Mm) and the outer penumbra ($r=13-15$ Mm) as indicated by the cyan circles in Figure \ref{fig:1}. Green dashed lines indicate the averaged $\tau=1$ and $\tau=0.01$ levels. The purple curve is an azimuthally and temporally averaged iso-contour placed at $B=950$ G and used as reference to distinguish between the high- and low-field regions in the penumbra. White contour lines on the top plots enclose regions with vertical upflow speeds larger than 0.3 km  s$^{-1}$ when using masks that select only the sources of inflows (i.e. grid-points where $v_r<0$ and $v_z>0$, left) and the sources of outflows (i.e. grid-points where $v_r>0$ and $v_z>0$, right).}\label{fig:8}
\end{figure*}

\begin{figure*}[t]
    \centering

    \includegraphics[width=0.45\textwidth]{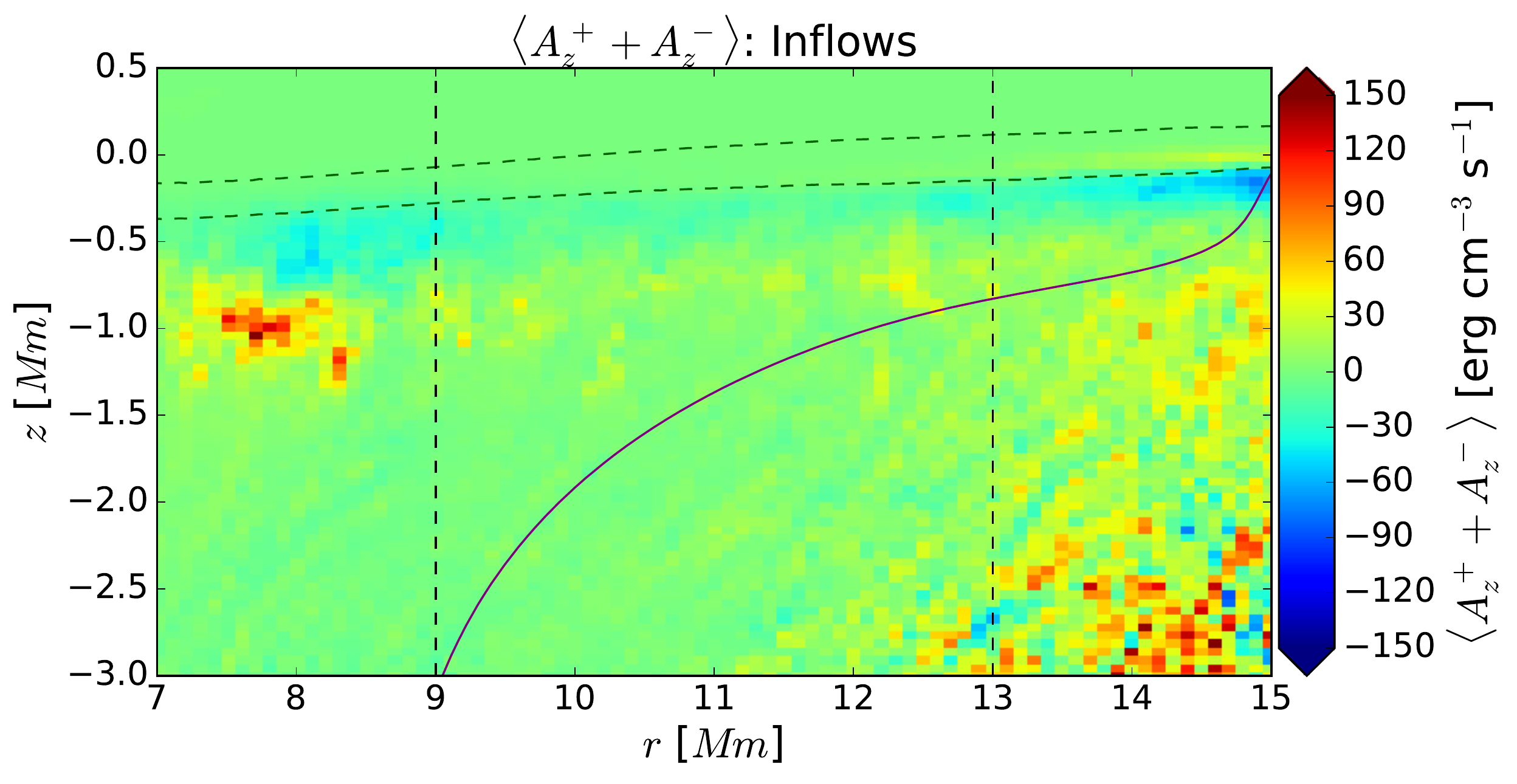} 
	  \includegraphics[width=0.45\textwidth]{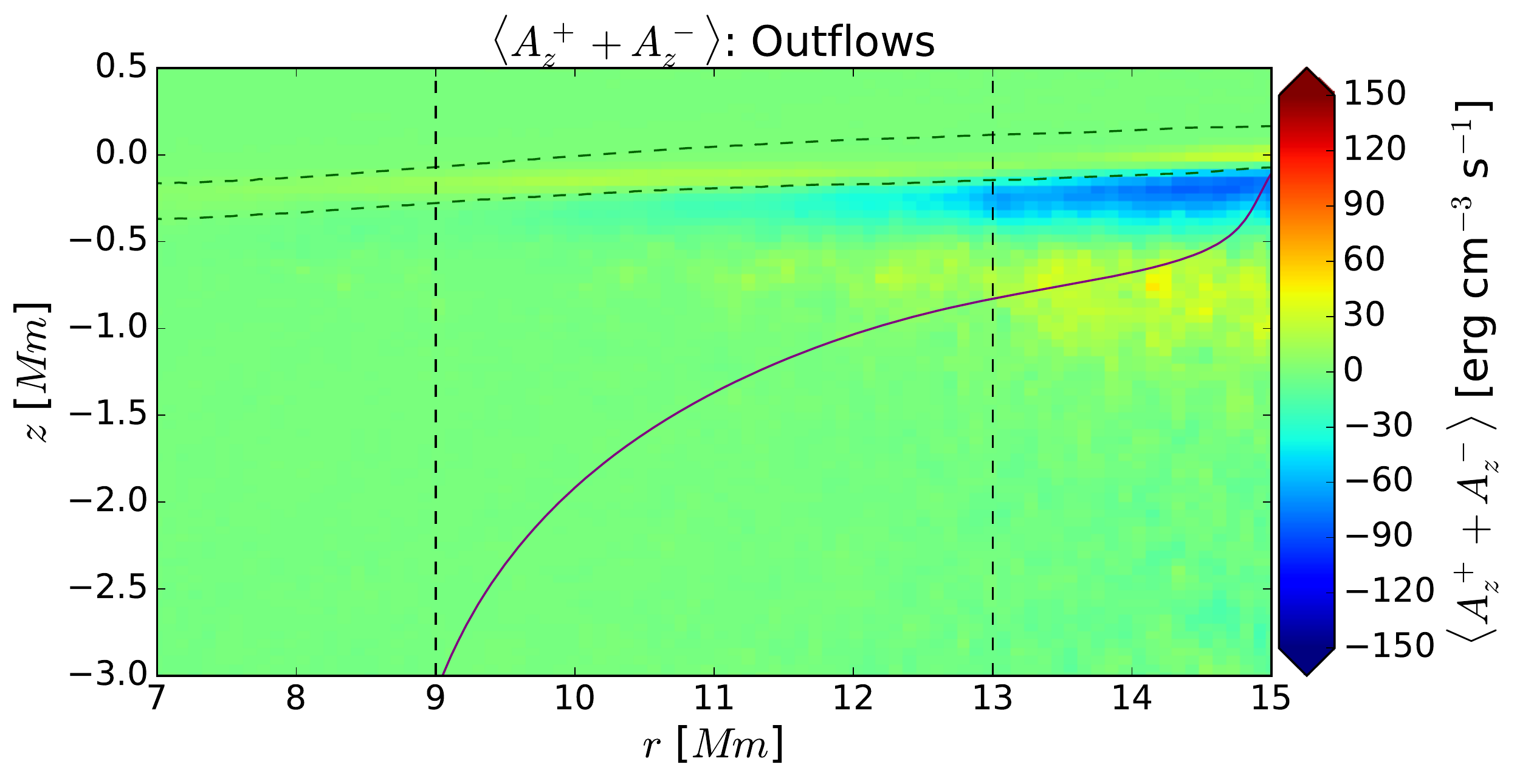} 
\includegraphics[width=0.45\textwidth]{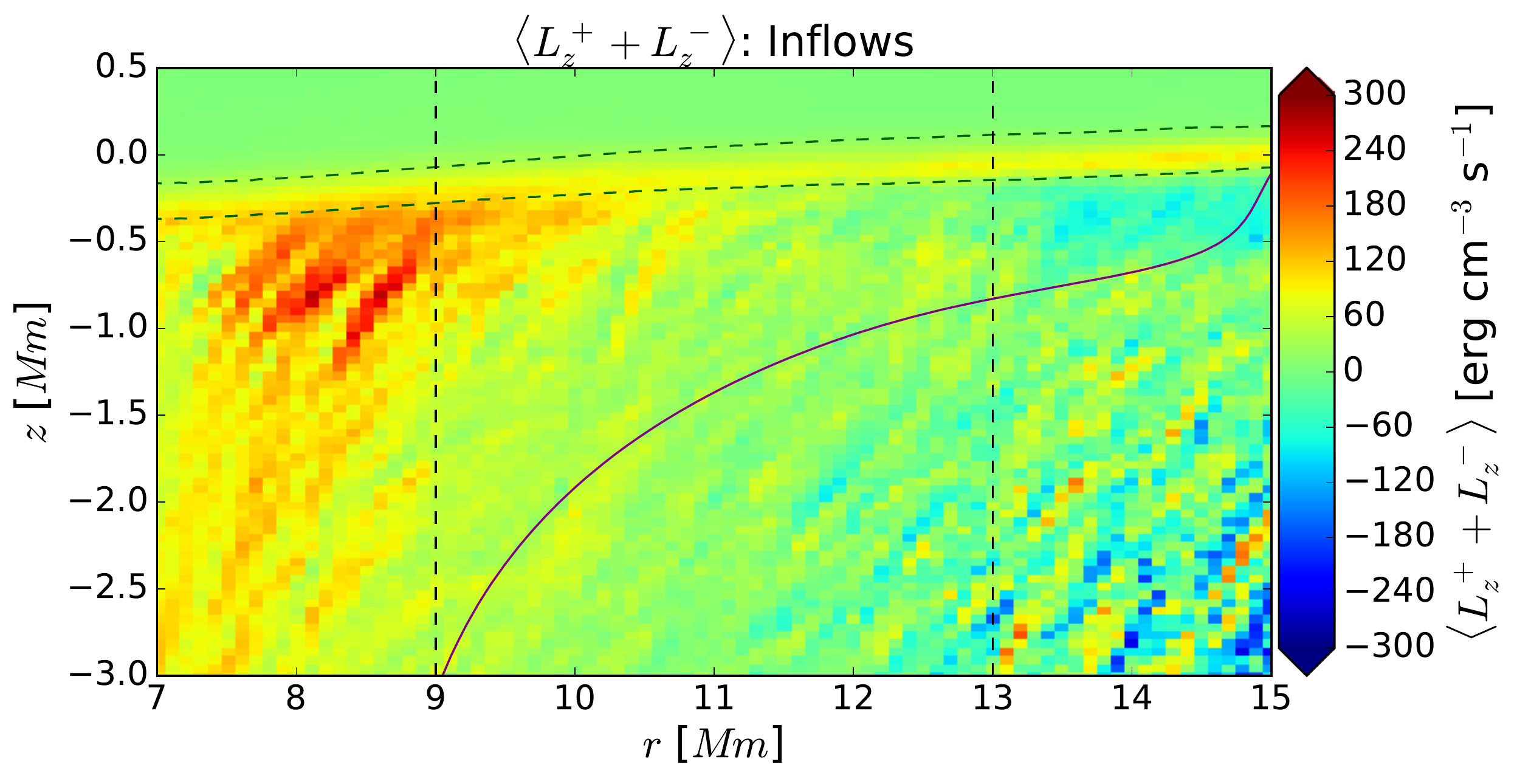} 
	  \includegraphics[width=0.45\textwidth]{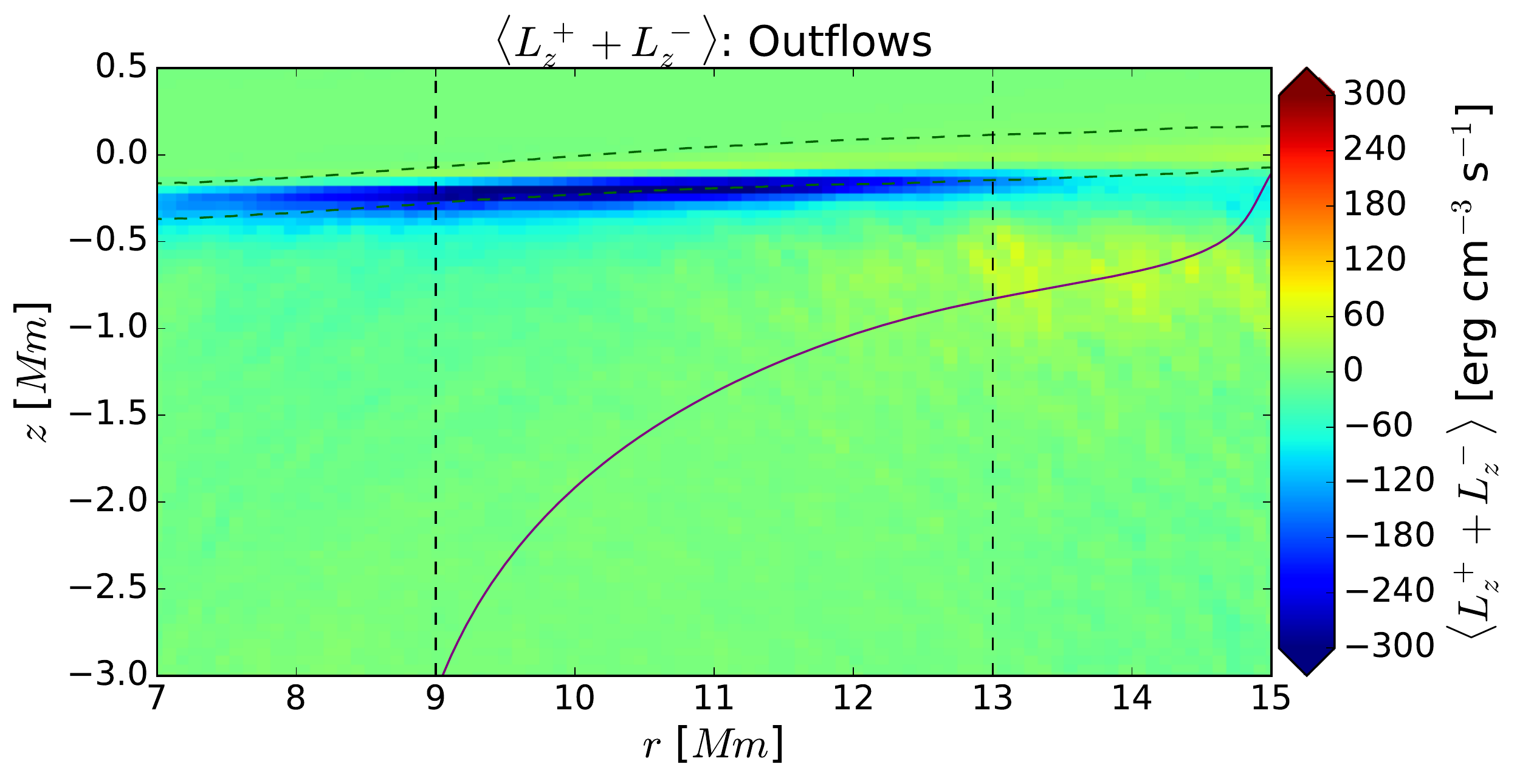} 
\includegraphics[width=0.45\textwidth]{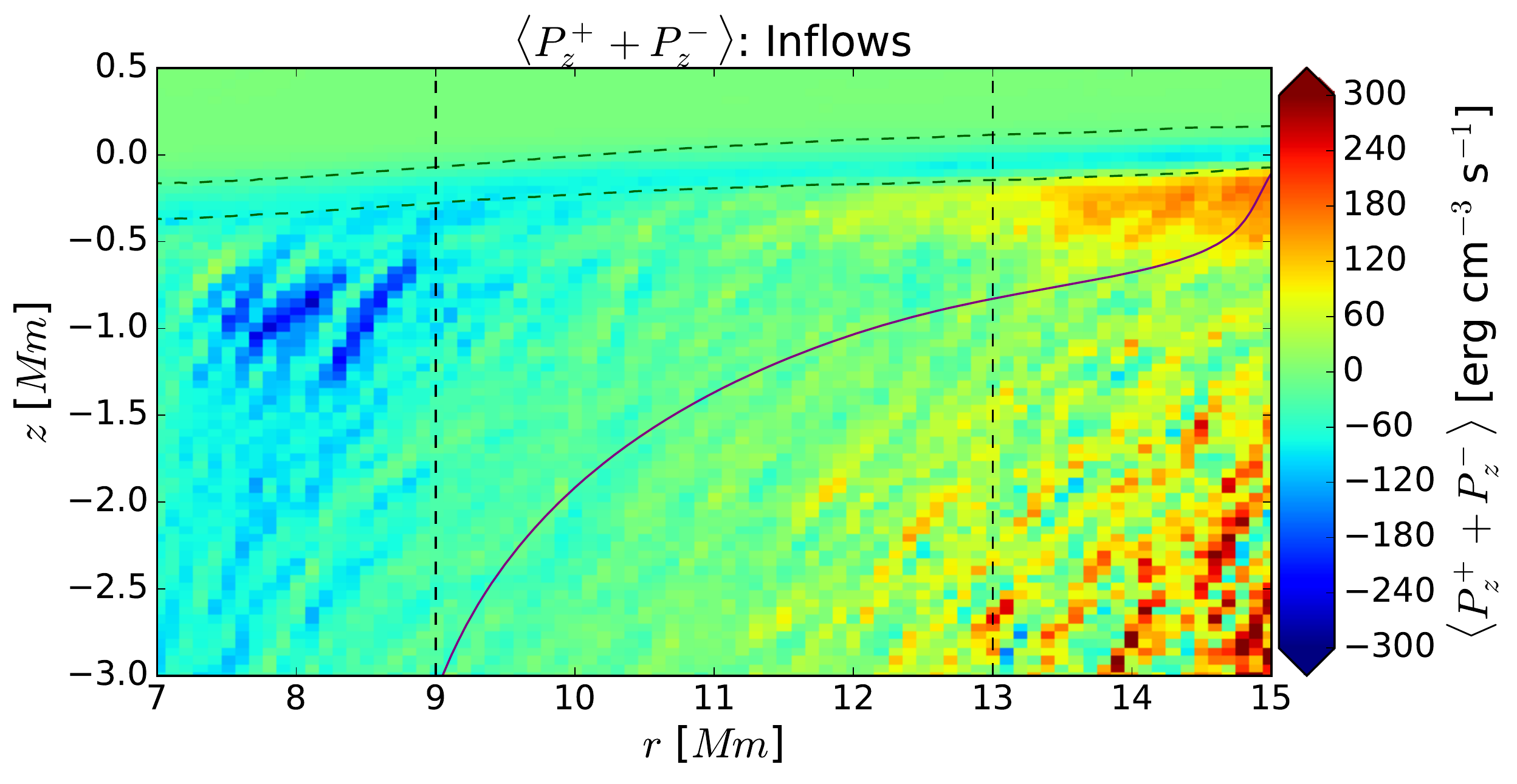} 
	  \includegraphics[width=0.45\textwidth]{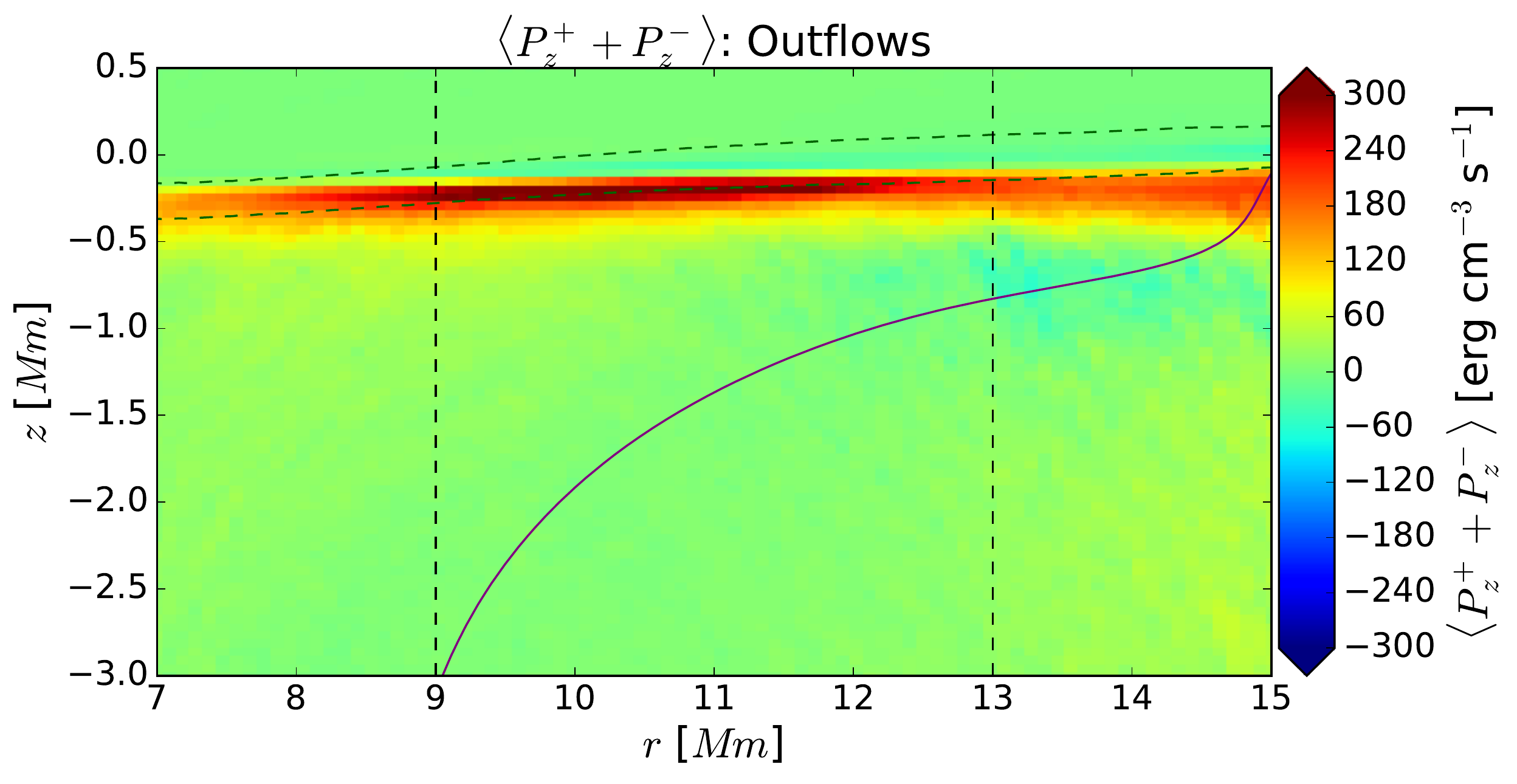} 
    \caption{Vertical energy conversion terms as functions of radius and height in the penumbra (azimuthal and temporal averages). From top to bottom: $\langle A_z^+ + A_z^- \rangle$, $\langle L_z^+ + L_z^- \rangle$, and  $\langle P_z^+ + P_z^- \rangle$ in  inflow regions (panels on the left) and in outflow regions (panels on the right). Same format as plots in Figure \ref{fig:8}.}\label{fig:8*}
\end{figure*}

\begin{figure*}[t]
    \centering
	 
\includegraphics[width=0.45\textwidth]{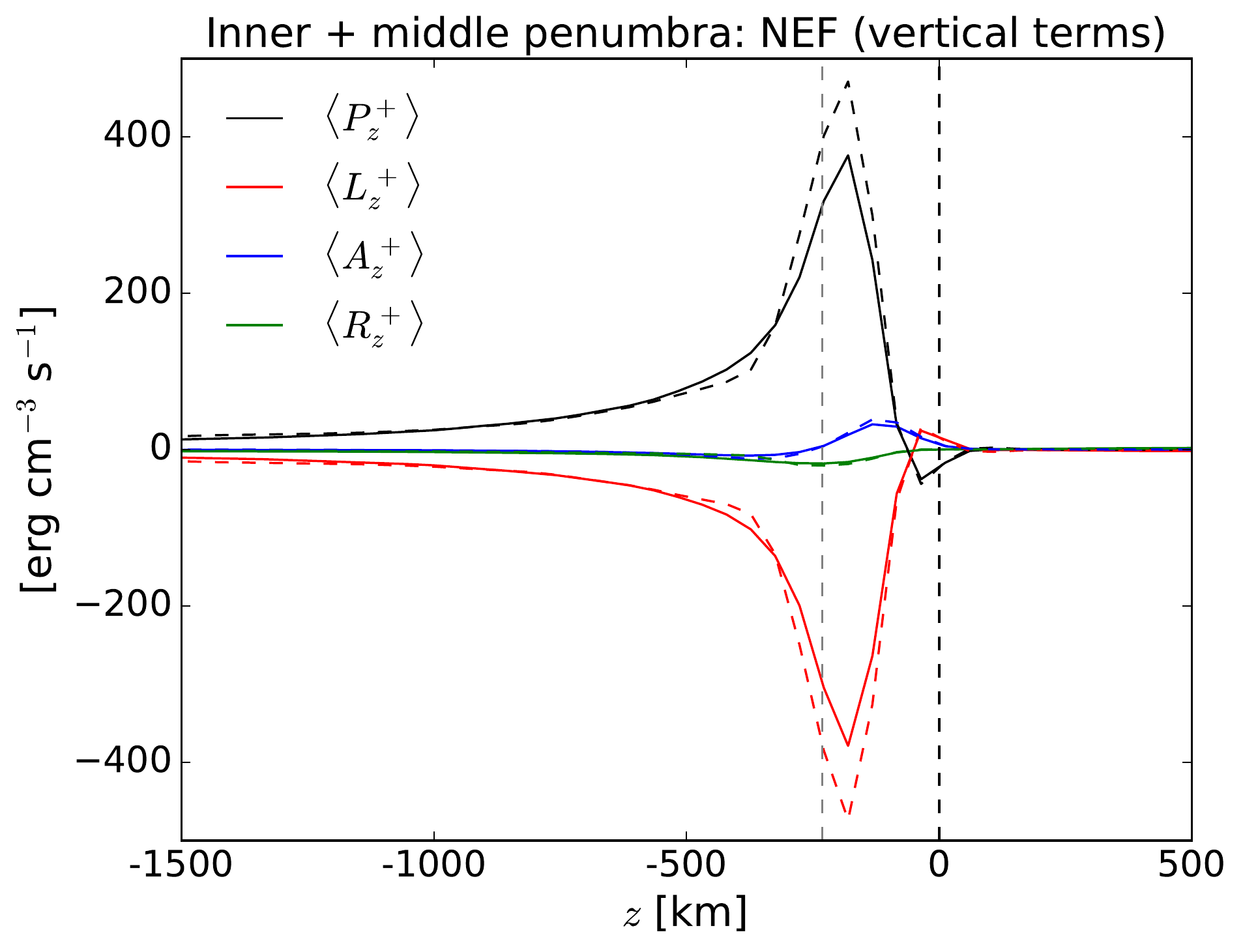} 
\includegraphics[width=0.45\textwidth]{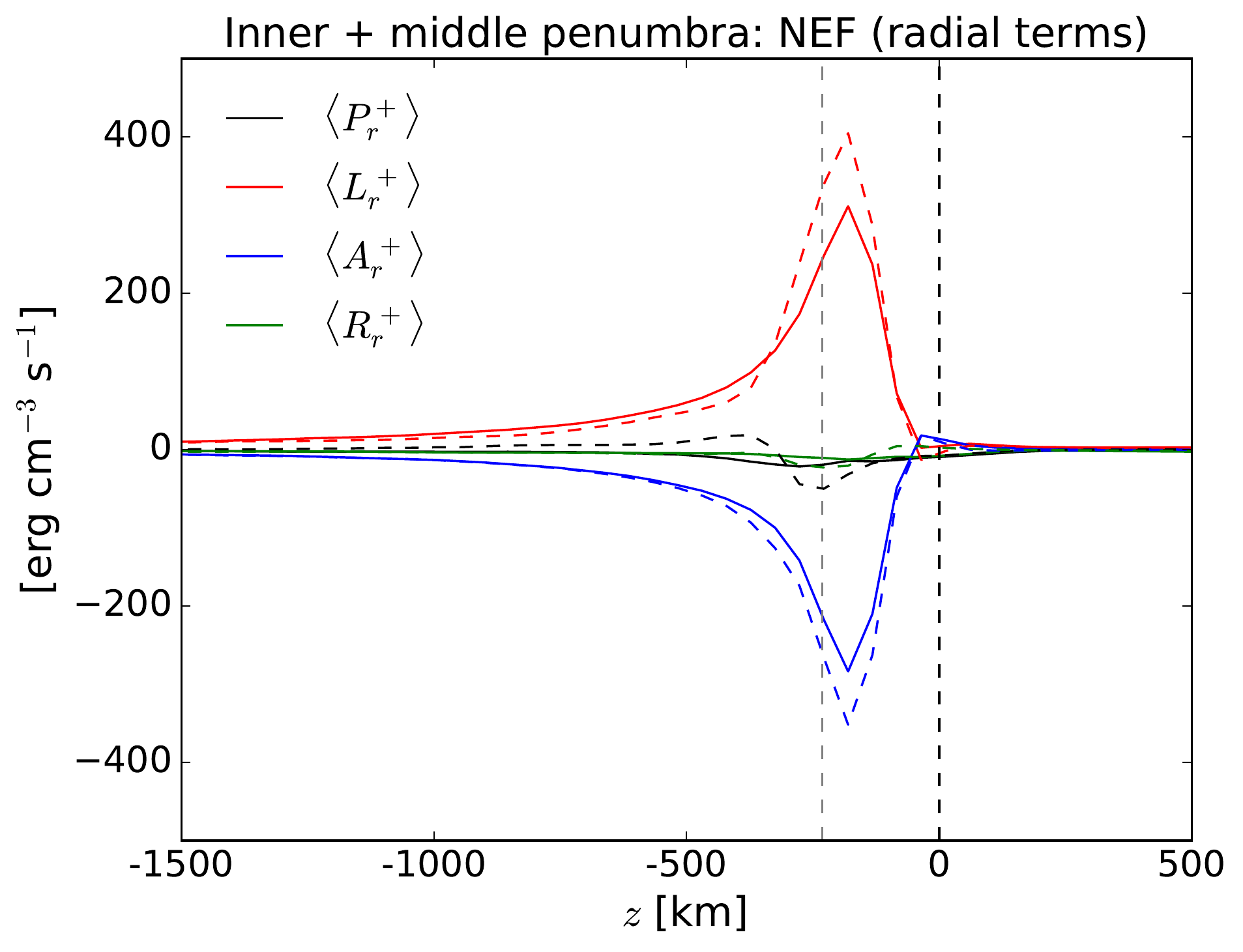} 
 \includegraphics[width=0.45\textwidth]{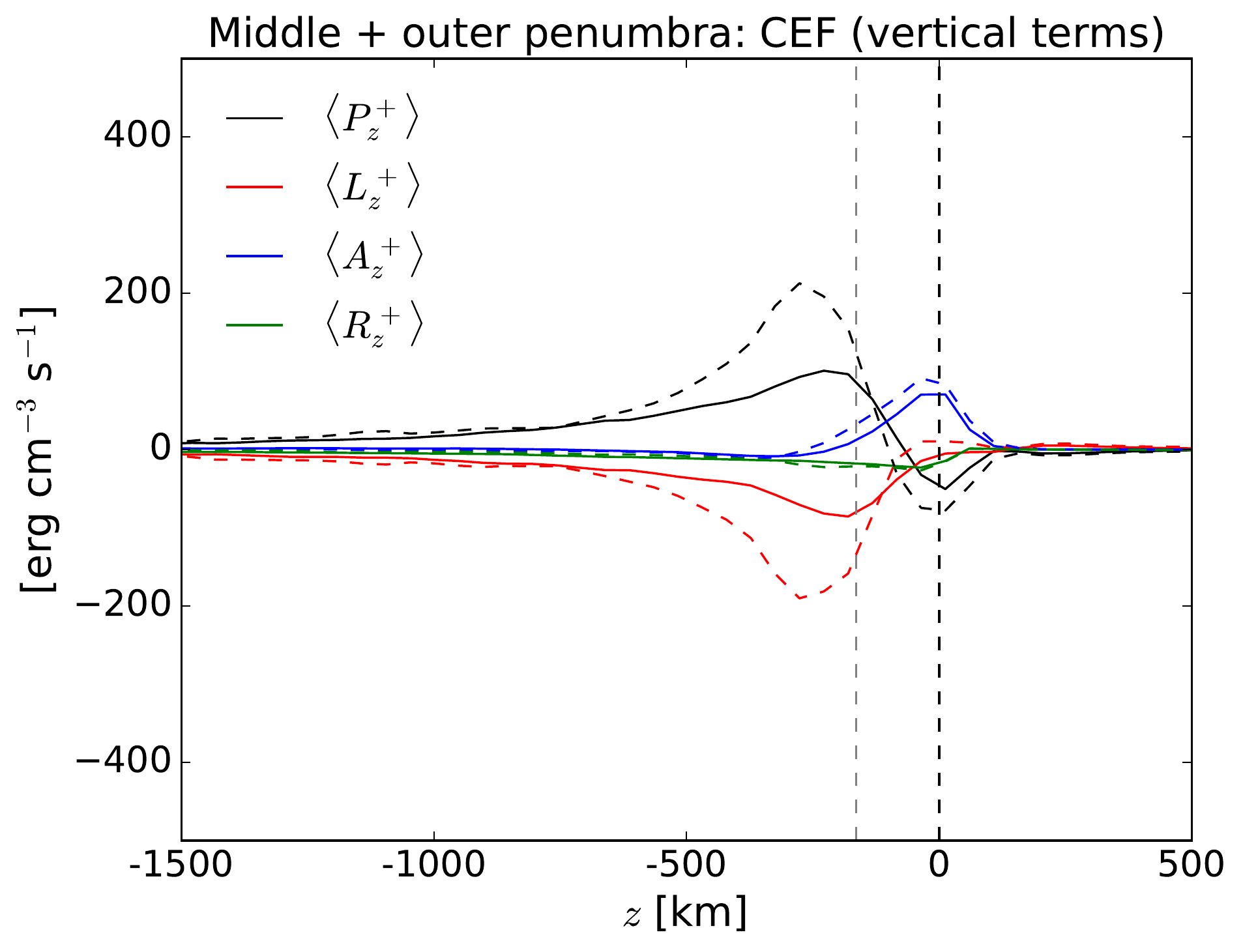} 
\includegraphics[width=0.45\textwidth]{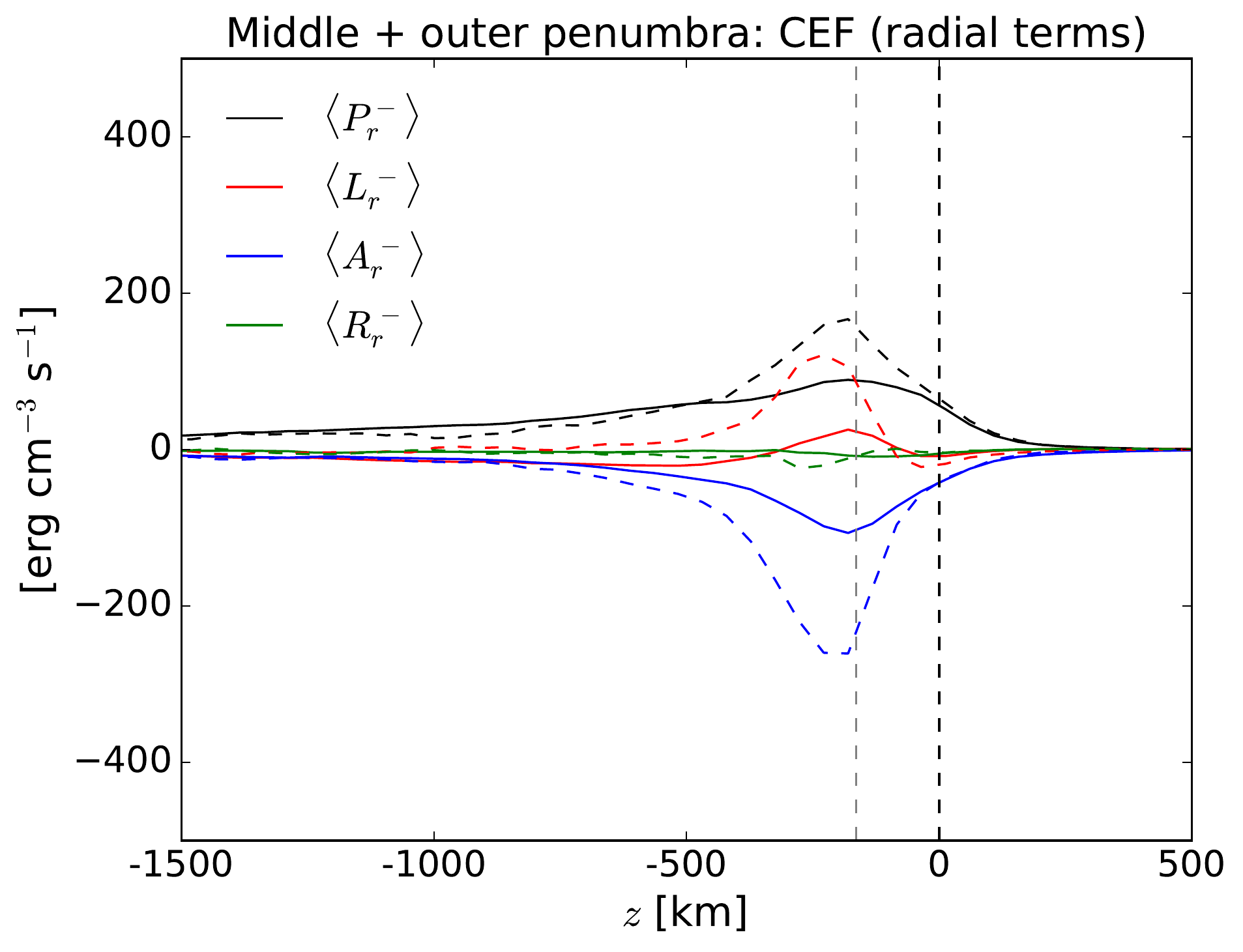} 
    \caption{ Height dependence of the energy conversion terms in the kinetic energy equation. The solid curves show the energy terms  averaged in time (from 60 to 70 hr) and at each height  for the regions that have both upflows and outflows ($v_z>0$ and $v_r>0$, upper plots) and for the regions with both, upflows and  inflows ($v_z>0$ and $v_r<0$, lower plots).
The dashed curves show the averaged energy terms when
using a mask that extracts the penumbral filaments carrying the CEF and the NEF, correspondingly, and neglects the contribution of 
 all other flows (e.g. those occurring beyond the reference iso-contour at $B=950$ G shown in Figures \ref{fig:8} and \ref{fig:8*}). 
 We considered only flows  in the inner and middle penumbra for the plots on the top, and in the middle and outer penumbra for the plots on the bottom. The energy conversion terms are separated into their vertical components (left plots) and radial components (right plots). Black: pressure forces, red:  Lorentz forces, blue: acceleration forces, green: residual forces. 
The vertical dashed lines are placed at  $z=0$ km  (black) and at the average height of the $\tau=1$ level in the corresponding regime(s) of the penumbra, which are indicated at the headers of each plot. }\label{fig:6}
\end{figure*}

In the equations \ref{eq:2}-\ref{eq:5}, $i$ indicates the Cartesian coordinates and the brackets indicate temporal averages over 10 hours, from $t=60-70$ hours after initialization of the simulation. We then use the transformation to cylindrical coordinates to separate the directions along and perpendicular to the penumbral filaments, $r$ and $z$ coordinates, respectively. With $v_i^{\pm}$ denoting positive and negative velocity components respectively ($v_r^+$ represents outflows and $v_z^+$ upflows).

The top panels of Figure \ref{fig:8} show the azimuthally and temporally averaged vertical slices of the  radial velocity separately for all the inflows (left) and all the outflows (right) in the penumbra, i.e. using masks that select only the grid points where $v_r<0$ and $v_r>0$ respectively, from $z=0.5$ to $-3$ Mm (relative to the average height of the $\tau=1$ level in the quiet Sun). 
The NEF-related outflow speeds larger than $2$ km s$^{-1}$, and in some places even larger than $5$ km s$^{-1}$ stand out close to the $\tau=1$ level, being generally restricted to  below $\tau=0.01$. Note that in this case the mask excludes any
inflow occurring in the penumbra, even those corresponding to the IEF-like feature which dominates the penumbra above $\tau=0.01$  (see radial velocity maps in Figure \ref{fig:2}). In contrast, 
 the inflow  speeds above 2 km s$^{-1}$, which  also stand out close to the $\tau=1$ level due to the CEFs, produce a continuously increasing inflow profile towards the higher layers due to the inclusion of all those inflows in the IEF-like feature that peak above $\tau=0.01$ .  It is not in the scope of the present work to analyze such an IEF-like feature. Moreover,  given that these inflows occur close to the top boundary of the simulation box, which is numerically closed, they likely are highly influenced by the upper boundary conditions.

Figure \ref{fig:8} also shows  the resultant radial energy conversion terms 
in the penumbra, separated for all inflows ($\langle A_r^- \rangle$, $\langle L_r^- \rangle$ and $\langle P_r^- \rangle$, plots on the left) and for all outflows  ($\langle A_r^+ \rangle$, $\langle L_r^+ \rangle$ and $\langle P_r^+ \rangle$, plots on the right). Recall that these energy conversion profiles show the 10  hours average of the azimuthally averaged terms in each mask. 
Negative values of $A_r^-$ and  $A_r^+$ indicate inward and outward acceleration of the fluid, respectively. Likewise,  positive values of $P_r^{-}$ and $L_r^{-}$ represent a source of kinetic energy for inflows,
whilst positive values of $P_r^{+}$ and $L_r^{+}$ represent an energy source for outflows.

Panels in Figure \ref{fig:8*} are in the same format as panels in Figure \ref{fig:8} and show the vertical energy conversion terms ($\langle A_z^+ + A_z^- \rangle$, $\langle L_z^+ + L_z^- \rangle$ and $\langle P_z^+ + P_z^- \rangle$) associated to the inflows (panels on the left) and to the outflows (panels on the right). There, negative values of $ A_z^+ + A_z^- $  indicate vertical acceleration of the fluid in inflow and outflow regions, accordingly. 

 While most of the radial acceleration of the fluid  takes place within  the inner and middle penumbra  for the outflows (negative values of $\langle A_r^+ \rangle$ in Figure \ref{fig:8}), the inflows are predominantly accelerated within the middle and outer penumbra (negative values of $\langle A_r^- \rangle$). 

These regions are confined to a narrow layer near the $\tau=1$ level. There, the Lorentz force is the primary driver of the outflows, while the pressure terms have a weakly negative energetic contribution. In contrast, the inflows are driven mainly by the radial pressure forces, while the radial Lorentz force shows a negative contribution on average. 
However, plots in Figure \ref{fig:6} reveal that the radial Lorentz term contributes mostly positively close to $\tau=1$ when the masks only include the regions responsible for driving inflows (i.e. source regions where both $v_r<0$ and $v_z>0$), and neglect those regions where the inflows sink (i.e. regions where $v_r<0$ and $v_z\leq0$).

Plots in Figure \ref{fig:8} also show that in deeper layers the radial pressure terms are almost in balance with the radial Lorentz terms, 
resulting in only minor acceleration of outflows and inflows in the deep penumbra. Likewise, almost no driving forces exist above $\tau=0.01$.
Towards the inner penumbra for the inflows  and outer penumbra for the outflows, the layers below the $\tau=1$ level display a similar forcing pattern, though the energy conversion is on average larger at the inner penumbra for inflows than at the outer penumbra for outflows.
In both cases, from $\tau=1$ down to $z\sim-1$ Mm, the radial Lorentz  force is overcompensating the radial pressure driving, producing a radial deceleration of  inflows and outflows (positive values of $\langle A_r^- \rangle$ and $\langle A_r^+ \rangle$, respectively).
This energy is however transferred into a vertical acceleration of the fluid (negative values of $\langle A_z^+ + A_z^- \rangle$ in  Figure \ref{fig:8*}). 
Such vertical acceleration is mainly downwards at the inner penumbra for inflows and at the outer penumbra for outflows , i.e. negative values of $A_z^-$ dominate in the $\langle A_z^+ + A_z^- \rangle$ average profiles at those places.
Finally, the downflowing gas is decelerated  again close to   $z=-1$ Mm. This is seen as the change into positive values of $\langle A_z^+ + A_z^- \rangle$ in Figure \ref{fig:8*}.
In the case of the outflows, a transition towards "normal" overturning convection  (i.e. more granular-like upflows and downflows where vertical pressure forces lead mostly to vertical acceleration) starts taking place in the region $r>13$ Mm. This also contributes to the strong positive $\langle A_z^+ + A_z^- \rangle$ found right below $\tau=1$, which also coincides with positive $\langle P_z^+ + P_z^- \rangle$.

As mentioned above, most of the radial acceleration occurs from the  middle to the outer penumbra in inflows, and from the inner to the middle penumbra in outflows. Such penumbral regimes also correspond to the average location of the respective flow source regions (see white contour lines in Figure \ref{fig:8}).
Therefore, plots in Figure \ref{fig:6} (solid curves) have been created by using a mask that specifically selects the sources of the outflows (regions where $v_z>0$ and $v_r>0$)  within the inner and middle penumbra at each height, and the sources of the inflows (regions where $v_z>0$ and $v_r<0$) within the middle and outer penumbra at each height. The average depth profile of the energy conversion terms at the sources of each flow are then averaged in time (10 hours).
Plots in Figure \ref{fig:6} show that  most of the energy conversion takes place between $z=-500$ and 0 km, and represents the driving of the NEF and the CEF, correspondingly.

However, the solid curves in Figure \ref{fig:6} stand for the sources of everything that is flowing outwards (upper plots) and the sources of everything that is flowing inwards (lower plots) in the respective penumbral regimes. This includes the filaments of interest carrying the NEF and the CEF respectively, but there are also other and smaller scale flows that do enter the average.
The weight or contribution of such flows 
to the average energy conversion terms seems to be negligible below $z\sim-500$ km (by comparing with the dashed curves in Figure \ref{fig:6}) but is larger  in the near surface layers (between $z=-500$ km and $z=0$) where the driving of the NEF and the CEF occurs. However, those additional flows do not modify the average energy balance  of the NEF. Only in the radial energy balance of the CEF 
we see an increased contribution of the radial Lorentz term to the inward acceleration of the fluid when neglecting other inflows. This is because those additional inflows are mainly driven by radial pressure forces in regions where $L_r^-$ is generally null or slightly negative.

In both,  NEF and CEF sources, the upflow components show an approximate balance between $P_z^+$ and $L_z^+$, mainly below and close to $\tau=1$. 
Higher up,  the vertical pressure driving, which includes the contribution of gravity, changes sign and the $A_z^+$ becomes positive just below $z=0$, implying  upflows that decelerate  before reaching the  photosphere.
The  steepening of  $P_z^+$ in the near surface layers is caused by the presence of strong magnetic fields. However, the field strength is about a factor 2-3 weaker in the outer penumbra than in the inner penumbra, which explains why the energy conversion is lower at the sources of the CEF than at the sources of the NEF, 
and might be related to the transient and unstable aspects of the CEFs.

For the NEF, we find that  the energy extracted by the Lorentz force in the vertical direction, $L_z^+$, is almost completely transferred to the radial outward acceleration of the fluid, $A_r^+$, similar to the findings of \citet{Rempel2011b}. The energy conversion by radial Lorentz force, $L_r^+$, is mainly balanced by $A_r^+$, and $P_r^+$ has a slightly negative contribution close to $\tau=1$. 
This can be thought as a deflection of the vertical pressure driving by the Lorentz force ($v_z B_r\approx v_rB_z$), i.e., the magnetic field forms an almost 90 degree nozzle connecting the pressure driving in the vertical direction with a radial outward acceleration.  The Lorentz force does not do any net work.
The ultimate energy source is the pressure force due to a large vertical pressure gradient which steepens in the near surface layers.

In normal convection,  a loss of buoyancy of the gas near $\tau=1$ causes a deceleration of the upflows and, as the gas cannot go any higher, it is pushed horizontally to all sides by the rising gas underneath it (i.e. the decelerating gas builds up excess pressure which drives horizontal flows). In the NEF, there is also a loss of buoyancy of the gas near $\tau=1$, but in this case  the horizontal pressure gradients only play a role in the azimuthal direction (i.e. to drive the horizontal lateral flows) and have slightly negative contribution in the radial direction. For the NEF, it is the inclined magnetic field in the penumbra who plays the most important role. Because in the penumbra there is a magnetic field pointing mainly in one direction, the gas motion loses degrees of freedom and therefore it moves in a preferred direction (the radial one). But the field is not the main accelerating agent of the gas flow in the radial direction, i.e. the density of the gas decreases at the surface where it is deflected radially, 
and because of mass flux conservation, its speed must be higher. The gas accelerates outwards because more and more gas is coming from below all along the filament, i.e. mass conservation requires the gas to move faster and faster horizontally.
While in granulation the horizontally flowing gas can  carry mass away on the horizontal plane isotropically,  in the penumbra there are less degrees of freedom due to the magnetic structure, and therefore the horizontal flows must be faster in penumbra than in granules to carry away the same amount of mass (assuming equally strong upflows and equal horizontal size of upflowing patches). However, the existence of downflows at the sides of the filaments  also removes part of the horizontally flowing mass, therefore limiting the speed of the radial flows.

In contrast, the CEF is predominantly driven by radial pressure forces (and so are all inflows at all heights). There is an additional positive contribution from the radial Lorentz force close to $\tau=1$, and the combination of both (magnetic and pressure forces) leads  to the inward acceleration of the fluid. 
The role of the radial Lorentz term for driving inflows in the near surface layers becomes more important when we extract the CEF-carrying filaments only (dashed lines in Figure \ref{fig:6}). 

Similarly to the situation at the sources of the NEF,  at the sources of the CEF the upflowing gas experiences a transition from more vertical into horizontal field  due to a bending of the field occurring at the outer and middle penumbra (see e.g. the field inclination  along the CEF filament shown in Figure \ref{fig:4}). Thus,  at the sources of the CEF the upflowing gas 
is partly deflected inwards by the magnetic field (note that part of the energy extracted by the magnetic field in the vertical direction is transferred into the inward acceleration of the gas). However, unlike the outflows, the inflows are primarily driven by radial pressure forces making the CEF a strong candidate for a siphon flow, which is further supported by our analysis in Section 4.


\subsection{Temporal evolution of the CEF}

\begin{figure*}[t]
    \centering

    \includegraphics[width=0.8\textwidth]{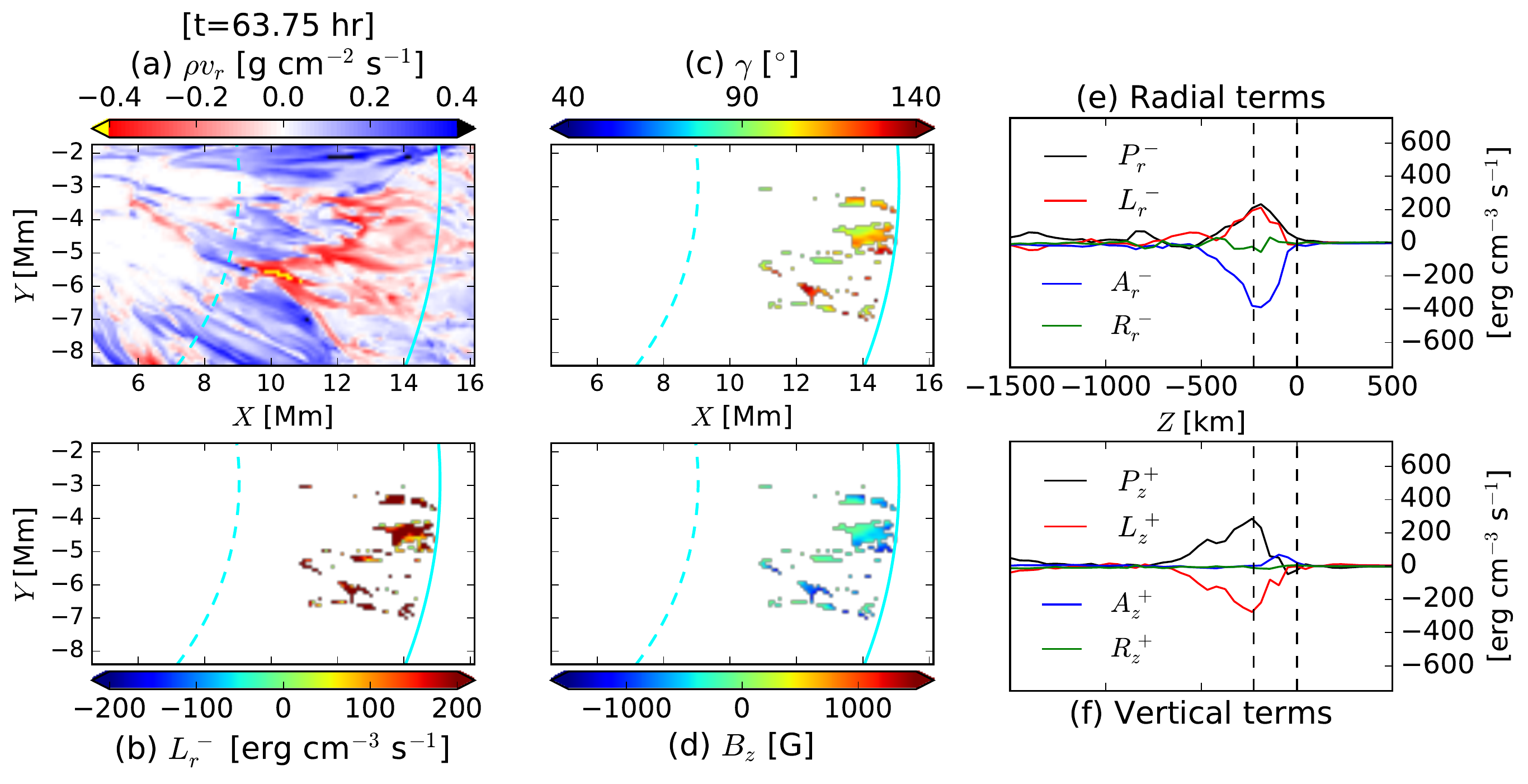} 
	  
\includegraphics[width=0.8\textwidth]{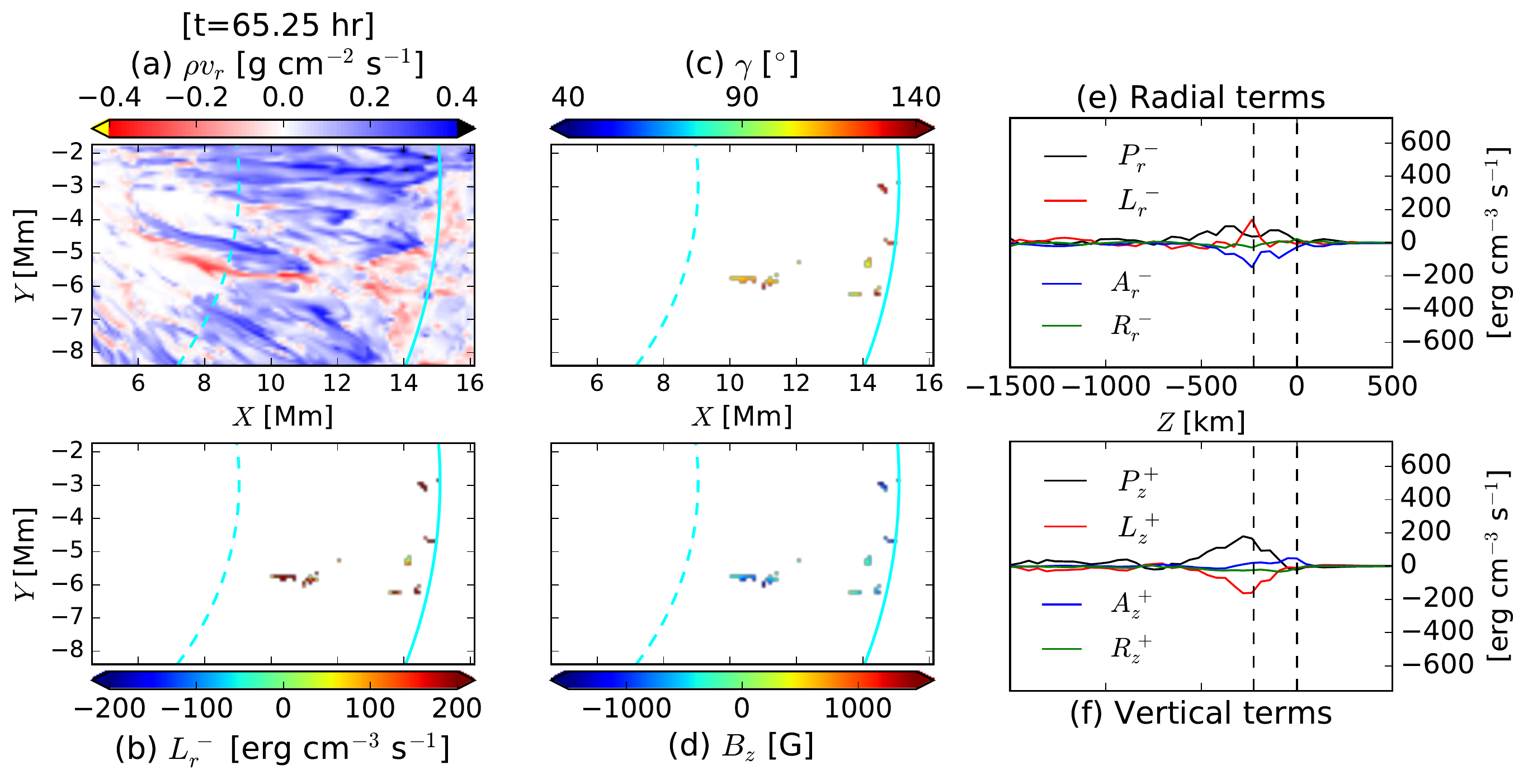} 
	  \includegraphics[width=0.8\textwidth]{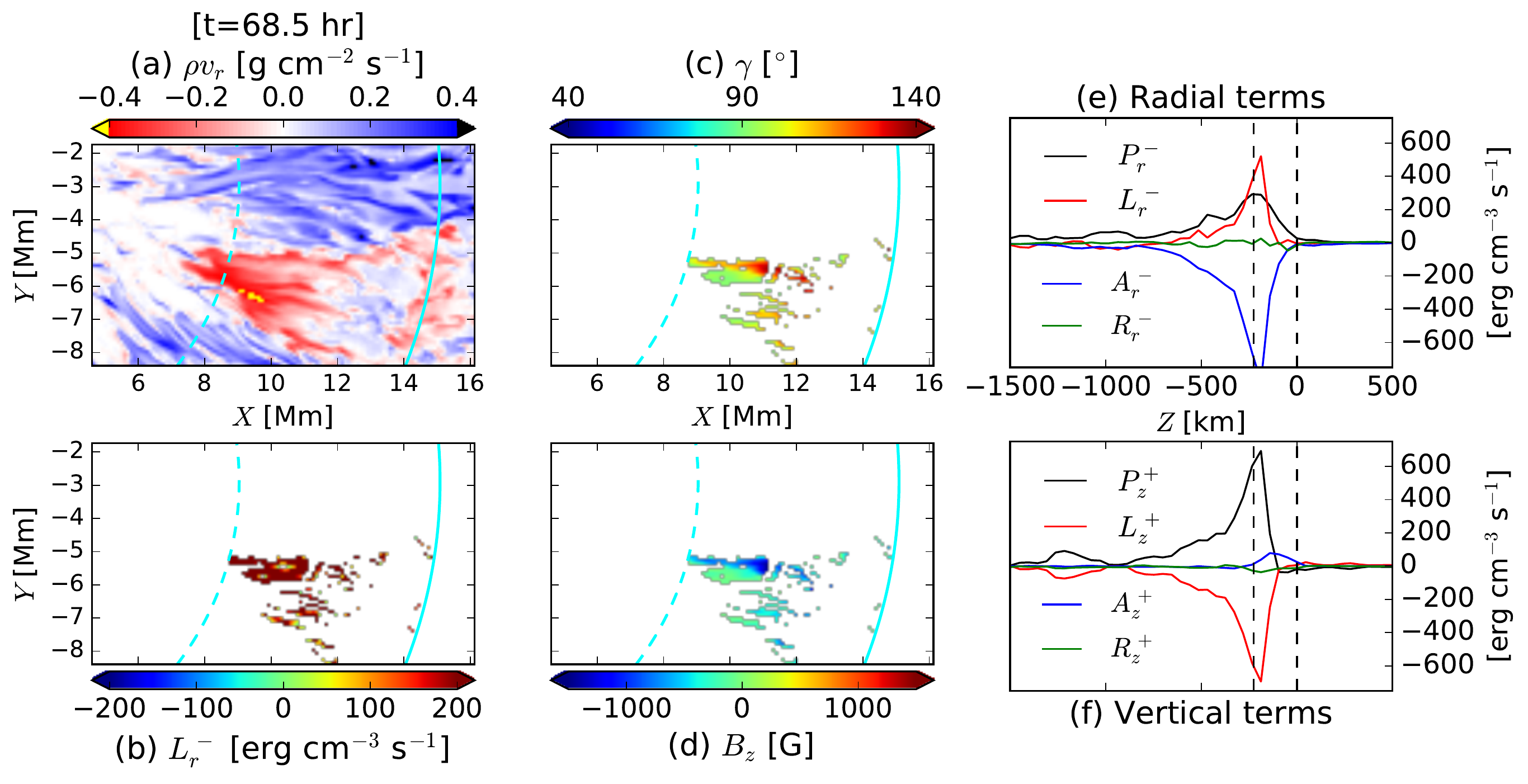} 
    \caption{Variability of the CEF in a portion of the penumbra at three selected time-steps, from top to bottom, $t=63.75, 65.25$ and 68.5 hours. For each time step six panels are plotted. Panel (a) displays the  radial mass flux, $\rho v_r$, at $z=-226$ km. Panels (b), (c) and (d)  show the quantities  $L_r^-$, field inclination $\gamma$, and vertical field component $B_z$, respectively, in the regions responsible for driving the CEF at $z=-226$ km, where the term $L_r^-$ has positive values. 
In panels (a)--(d) cyan curves were placed at $r=9$ Mm (dashed) and $r=15$ Mm.
Panels (e) and (f) depict the height profiles of the radial and the vertical energy conversion terms, respectively. The energy conversion terms have been averaged  over the sources of the CEF at each height and time-step, 
using the same mask as in the dashed plots of Figure \ref{fig:6}. The vertical dashed lines in panels (e) and (f) are placed at  $z=0$ km (average height of the $\tau=1$ level in the quiet sun) and at  $z=-226$ km (average height of the $\tau=1$ level in the penumbra).}\label{fig:9}
\end{figure*}

\begin{figure*}[t]
    \centering
    \includegraphics[width=0.3\textwidth]{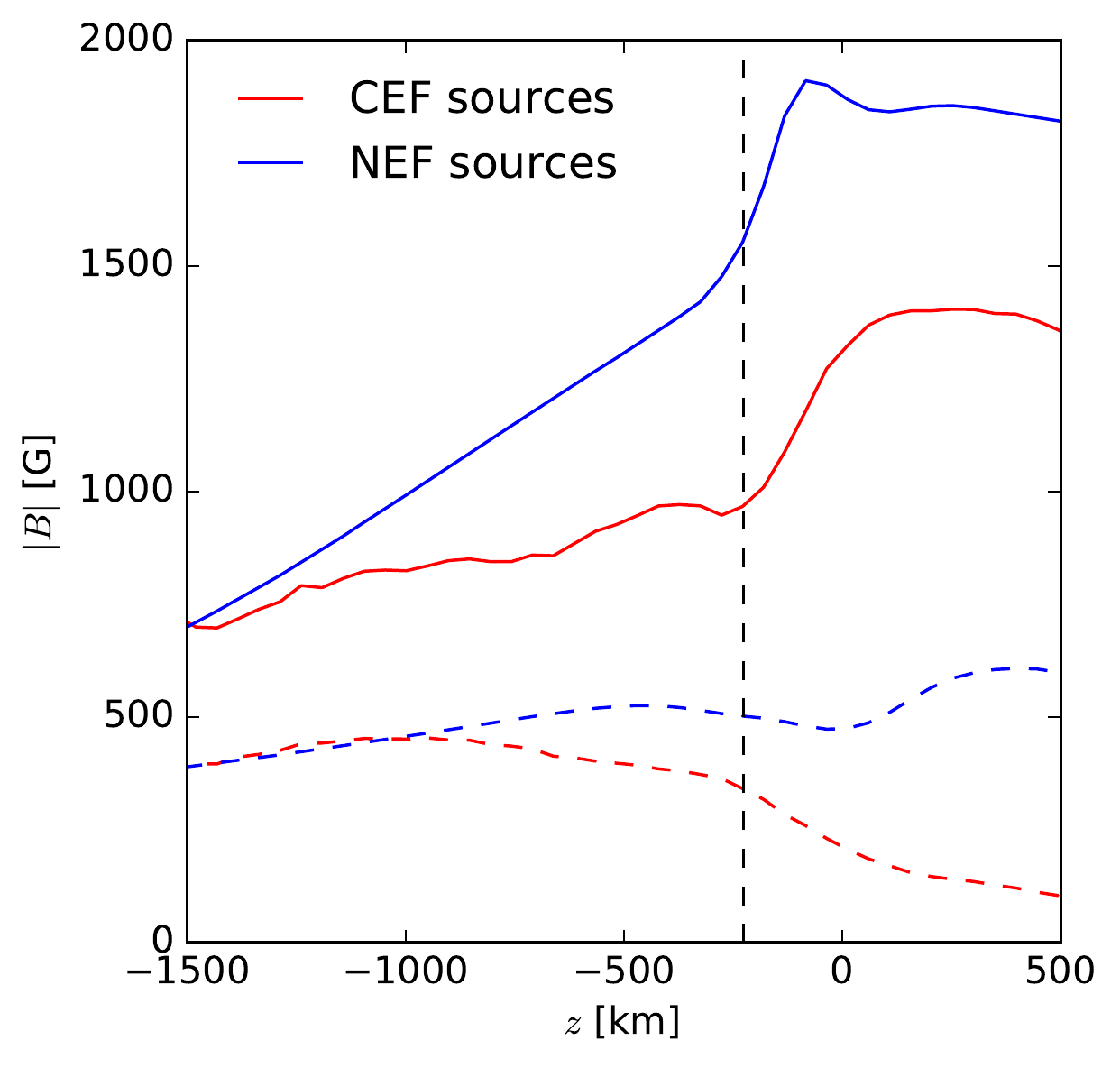}
 \includegraphics[width=0.3\textwidth]{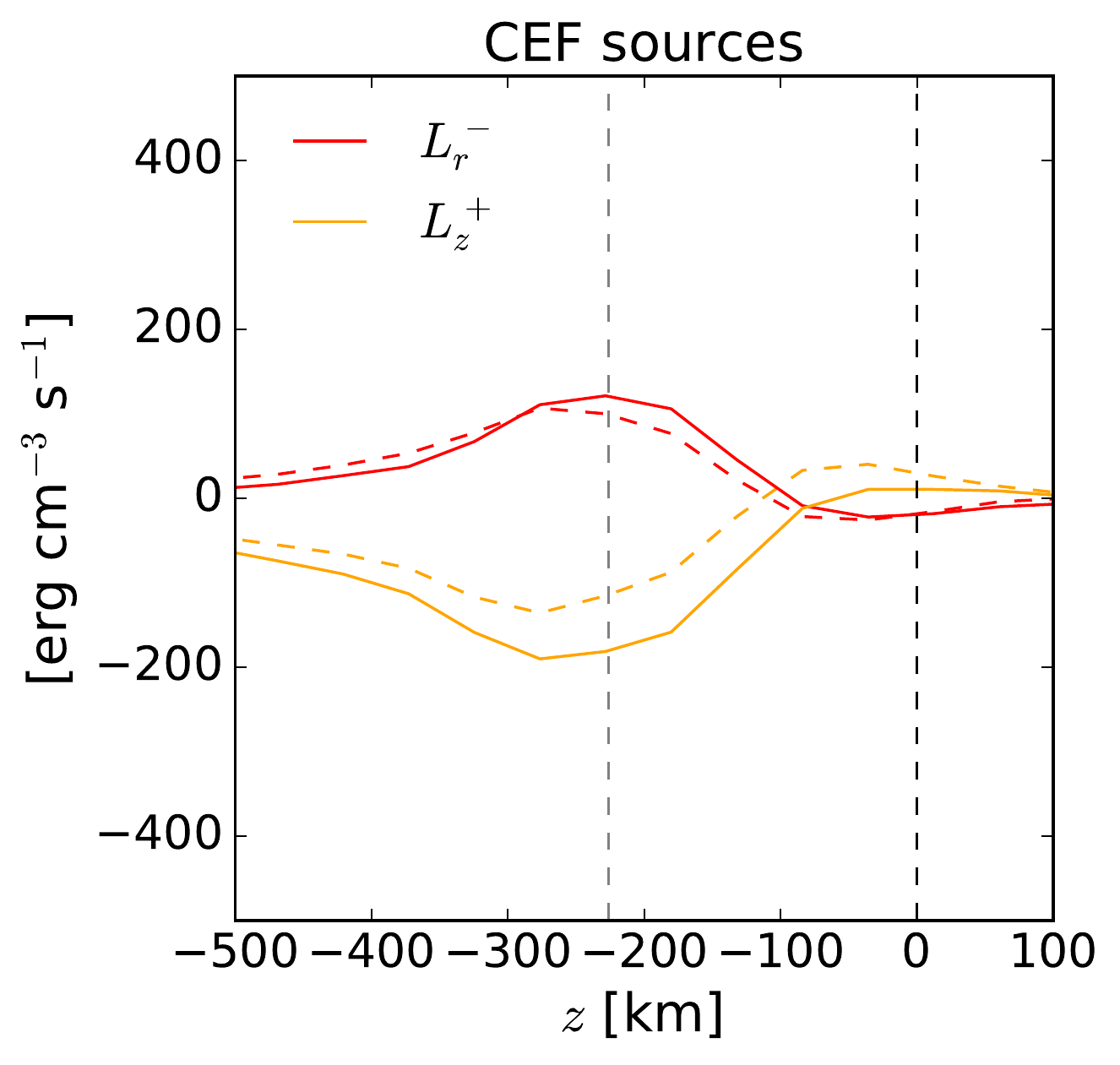}
 \includegraphics[width=0.3\textwidth]{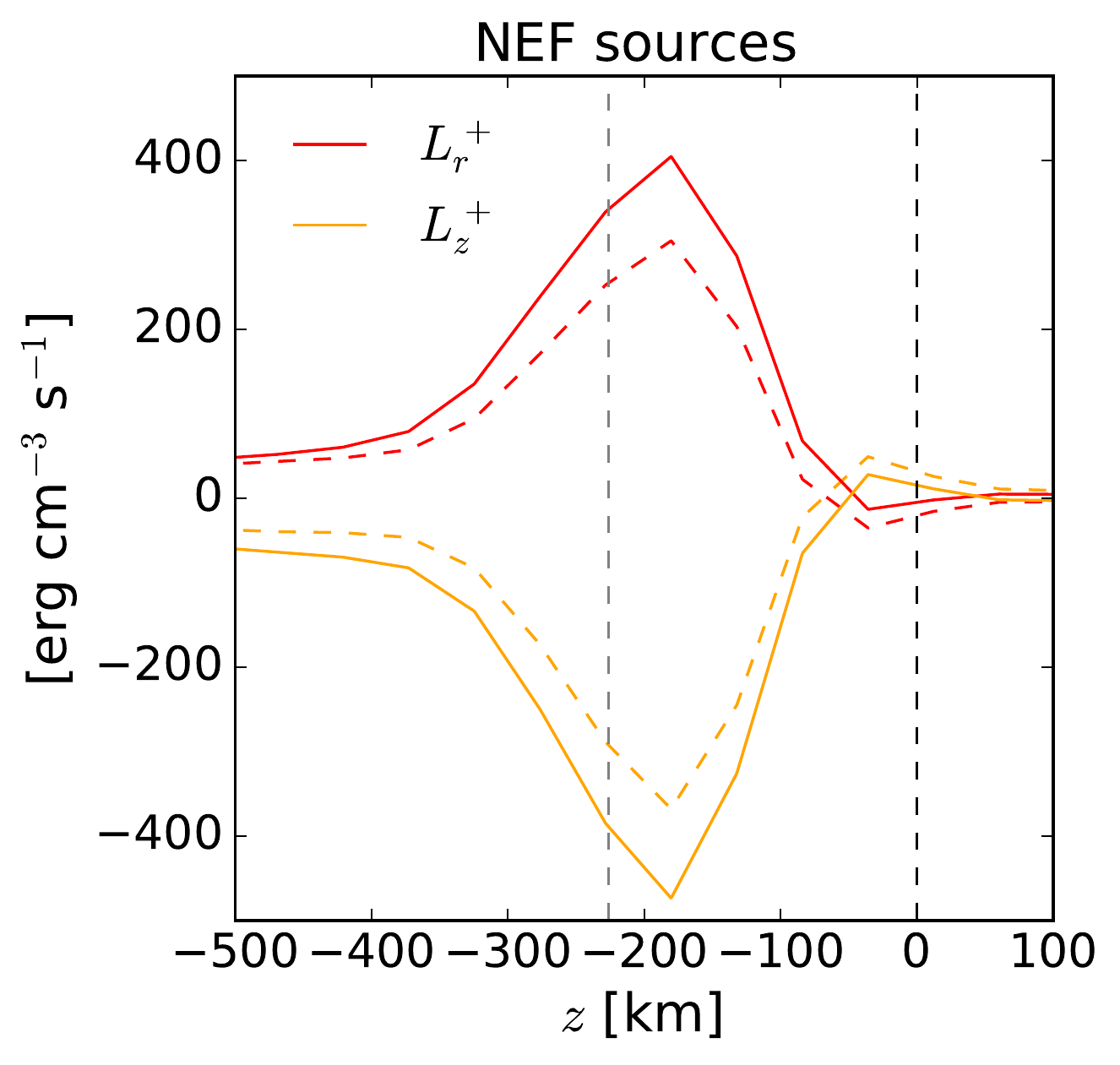}
	
    \caption{\textit{Left panel:} Temporal average  of the magnetic field vertical profiles in regions responsible for driving the CEF (red) and the NEF (blue). 
The magnetic field is separated according to its vertical (dashed) and radial (solid) components. The vertical dashed line is placed at $z=-226$ km.
\textit{Middle panel:} Energy conversion by radial (red) and vertical (orange) Lorentz force at the sources of the CEF. Solid lines are a close up view from the corresponding dashed lines in Figure \ref{fig:6}. The dashed lines show the simplified expression: $L_r=v_r\frac{1}{4\pi}B_z\frac{\partial B_r}{\partial z}$ (red) and $L_z=-v_z\frac{1}{4\pi}B_r\frac{\partial B_r}{\partial z}$ (orange). \textit{Right panel:} Same as in the middle panel for the sources of the NEF.}\label{fig:11*}
\end{figure*}

\begin{figure*}[t]
    \centering

  \includegraphics[width=0.48\textwidth]{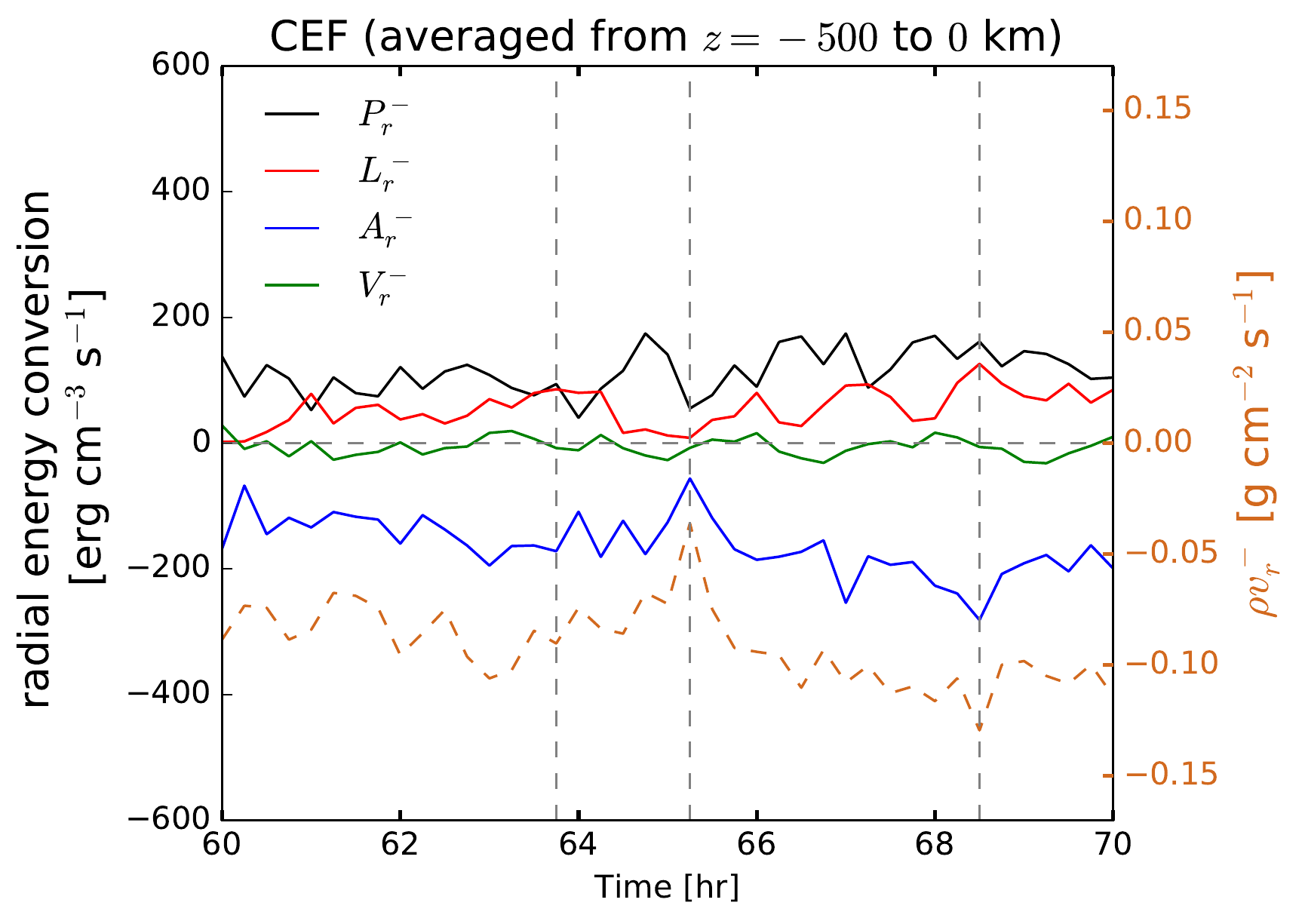} 
	  \includegraphics[width=0.48\textwidth]{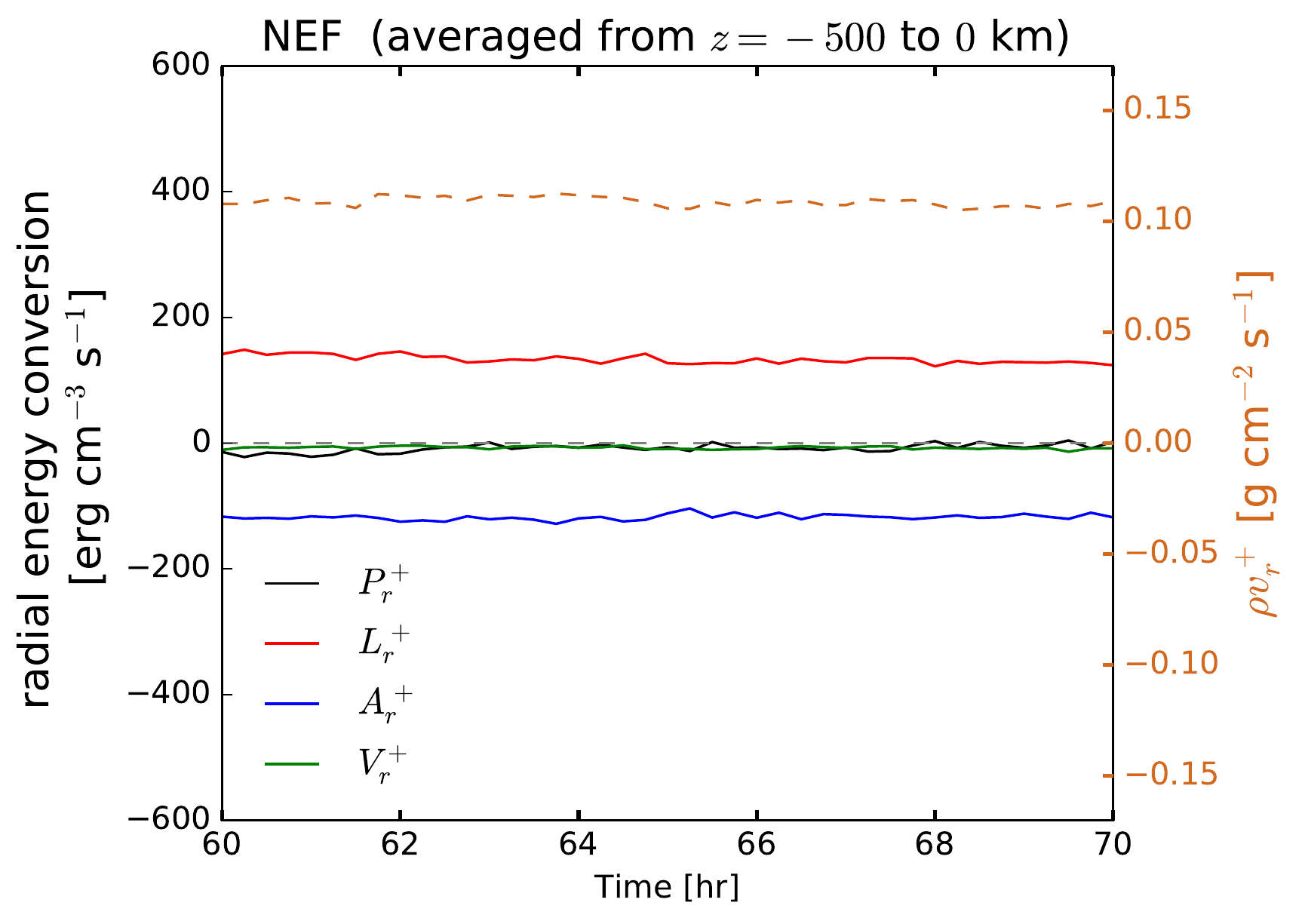} 
    \caption{Temporal evolution of the radial forces (solid) and mass fluxes (dashed) at the sources of the CEF within the portion of penumbra shown in Figure \ref{fig:9} (left plots). At each time, the plots show the average value of each quantity in the near-surface layers (between $z=-500$ and $z=0$ km) from the middle to the outer penumbra. 
For comparison, we show the same average quantities for the sources of the NEF (right plots) averaged over the inner to the middle penumbra. Vertical dashed lines on the left panel indicate the three selected stages of the CEF that are shown in Figure \ref{fig:9}.} \label{fig:10}
\end{figure*}

Figure \ref{fig:9} follows the evolution in time of a portion of the penumbra with CEF.
We focus on a single CEF patch only, since although the time evolution of the CEF patches in different parts of the penumbra are rather independent of each other, the general results described below are true for all the simulated CEF patches.

The figure shows in panels (a) the radial mass fluxes at three  stages of the evolution of the CEF in the selected penumbral sector at $z=-226$ km, the negative values indicate mass flowing radially inwards.
Panels (b), (c) and (d) in Figure \ref{fig:9} show that the source regions of the CEF at   $z=-226$ km that have positive values of $L_r^-$ are all characterized by negative values of $B_z$, i.e. downward pointing fields ($\gamma>90^{\circ}$) regardless of whether they are located in the middle or in the outer penumbra. In those regions,  the field inclination increases radially inwards.
Panels (e) show the radial energy conversion associated with the sources of the CEF in the different time-steps. The contribution of both, $L_r^-$ and $P_r^-$ to  the inward acceleration of the fluid vary over time, being the increase of the overall radial energy conversion associated with an increase of the radial mass flux in the selected CEF patch.
At all three stages of the CEF shown in Figure \ref{fig:9}, the vertical pressure force shown in panels (f) presents a steepening close to $z=-500$ km and it peaks near the average $\tau=1$ level of the penumbra ($z=-226$ km), keeping a close balance with work done against the Lorentz force.

In order to gain insight into the typical underlying magnetic structure of the regions responsible for driving the CEF, we show in Figure \ref{fig:11*} (left panel) the average 
vertical and radial magnetic field as a function of height in regions that are sources of CEF (red) and in regions that are sources of NEF (blue).
For the CEF sources,  the vertical field component (dashed red line) remains constant at about 400 G below $z=-226$ km and decreases to $\sim 100$ G in higher layers. The radial field component (solid red line) increases monotonically from about 700 G to $\sim1000$ G below $z=-226$ km. At  $z\sim-226$ km, the slope of $\vert B_r \vert$ suffers a large steepening and the radial field component increases to almost 1400 G towards $z=0$ and remains nearly constant at higher layers.
The steep increase of $\vert B_r \vert$ near $\tau=1$ combined with the reduction of $\vert B_z \vert$ results in a strong increase of the radial Lorentz force. As shown in the middle panel of Figure \ref{fig:11*}, there is a good agreement between the Lorentz terms as defined in equation \ref{eq:3} (solid lines) and the  following approximations:

\begin{equation}
L_z \approx -v_z \frac{1}{4\pi}B_r\frac{\partial B_r}{\partial z}
\end{equation} \label{eq:6}
\begin{equation}
L_r\approx v_r\frac{1}{4\pi}B_z\frac{\partial B_r}{\partial z}
\end{equation}\label{eq:7}

\noindent which are indicated by the dashed lines. Similar to what happens at the NEF sources (right panel), at the CEF sources  the energy extracted in the vertical direction by the Lorentz force, $L_z$, is in approximate balance with the radial Lorentz force, $L_r$. Then, for both, NEF and CEF sources, the following is valid:
\begin{equation}
v_z \frac{1}{4\pi}B_r\frac{\partial B_r}{\partial z}\approx v_r\frac{1}{4\pi}B_z\frac{\partial B_r}{\partial z}
\end{equation}

\noindent which leads to the relation $v_z B_r\approx v_rB_z$. This relation implies a  deflection of  the vertical pressure forces in upflows by 
the Lorentz force to allow for the strong acceleration in the radial 
direction and it is in agreement with the findings of \citet{Rempel2011b} for the driving mechanism of the NEF. Here, we show that this relation is also valid  for  the CEF, whose  vertical flows occur in regions where the vertical field component is negative and its magnitude decreasing with height, while the upflowing gas encounters an increased radial field component when it reaches the surface. This results in a radial driving by the Lorentz force that favors inflows and accelerates the gas inwards with the help of radial pressure forces, which in this case are larger than the Lorentz force.

Similarly, at the source regions of the NEF,  the average vertical field component (dashed blue line in left panel of Figure \ref{fig:11*}) remains around $\sim 400$ G whilst the radial field component (solid blue line) increases steadily from about 700 G to $\sim1500$ G just below $z=-226$ km. Just as at the CEF sources, the slope of $\vert B_r \vert$ suffers a strong steepening close to $z\sim-226$ km so that $\vert B_r \vert$ increases up to 1900 G towards $z=0$ and drops again in higher layers to about 1800 G.
Note that the average increase of $\vert B_r \vert$ near $\tau=1$ is substantially larger for the NEF sources than for the CEF sources because they occur within different parts of the penumbra. This causes  the contribution of the radial Lorentz term  to be larger at the sources of the NEF than at the sources of the CEF.

\begin{figure*}[t]
    \centering
\includegraphics[width=0.45\textwidth]{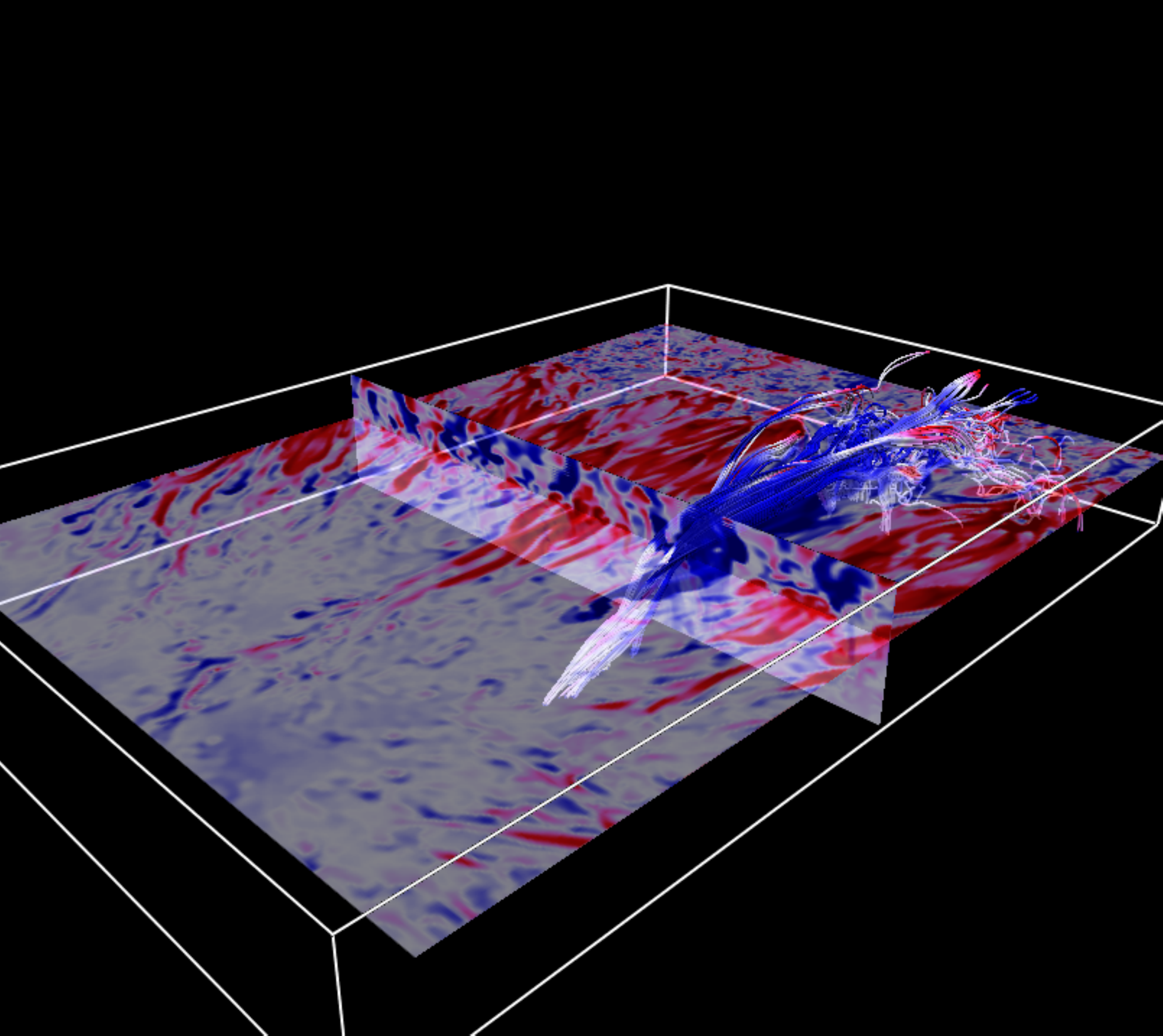} 
	 \includegraphics[width=0.45\textwidth]{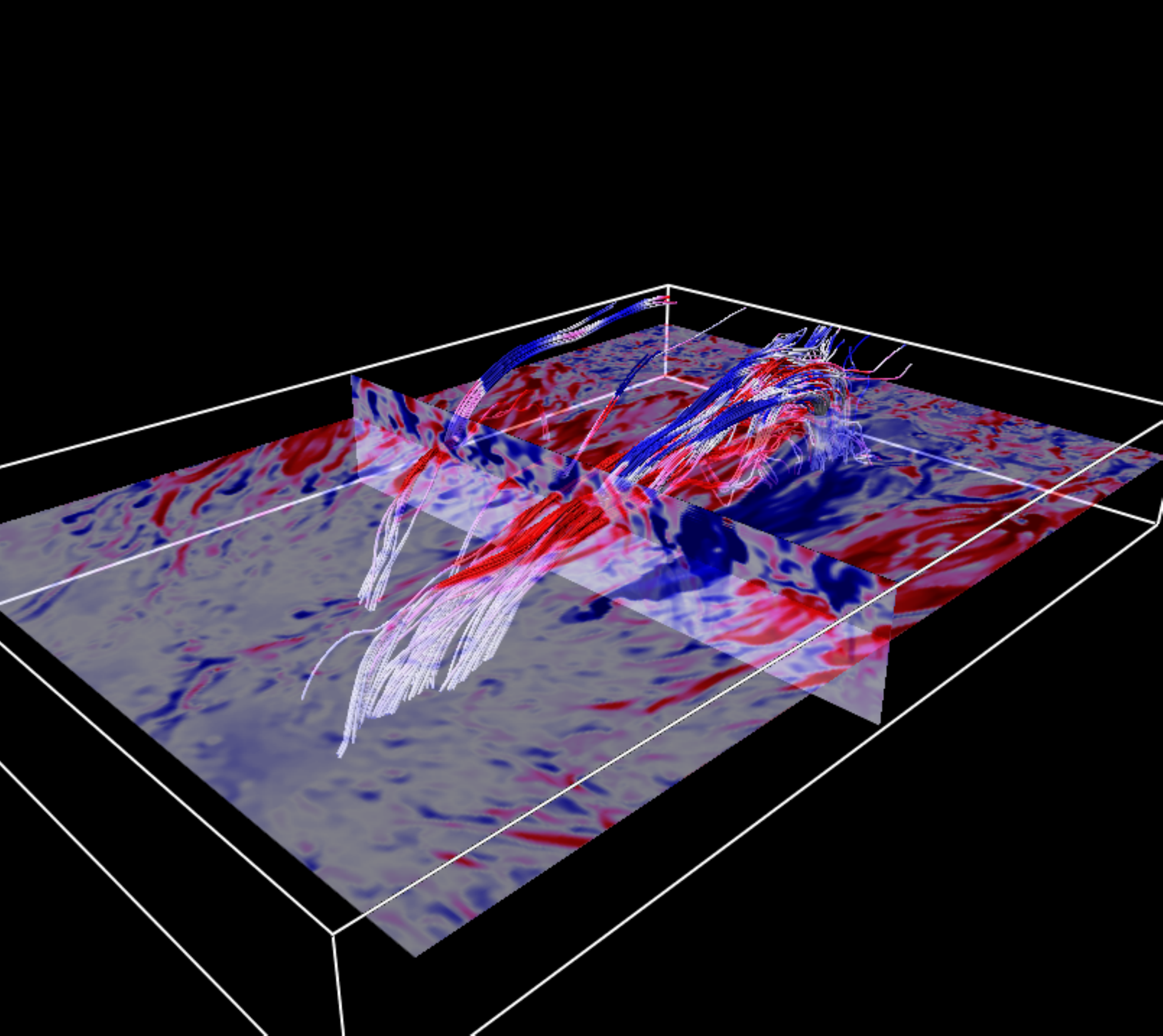} 
    \includegraphics[width=0.45\textwidth]{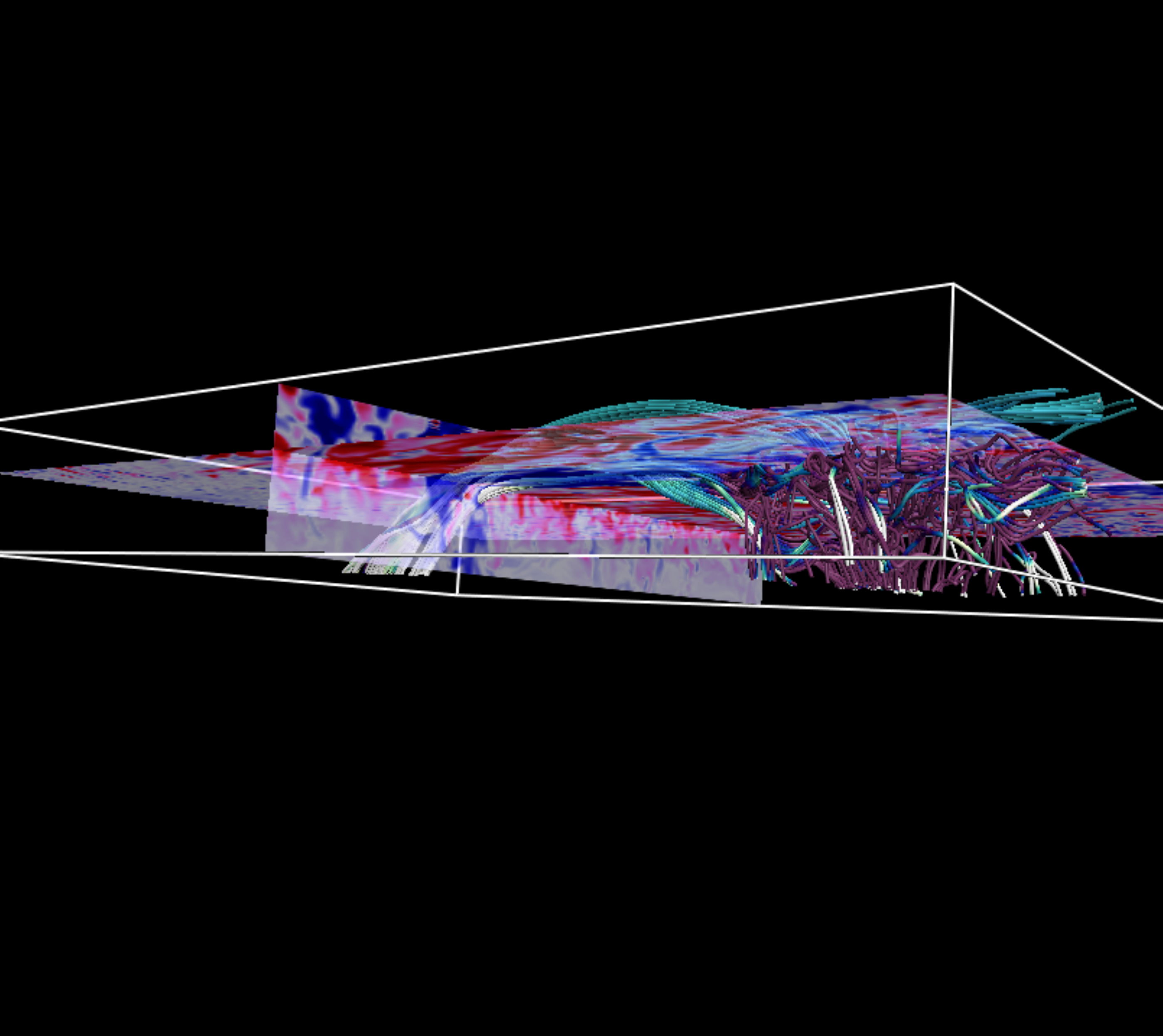} 
	 \includegraphics[width=0.45\textwidth]{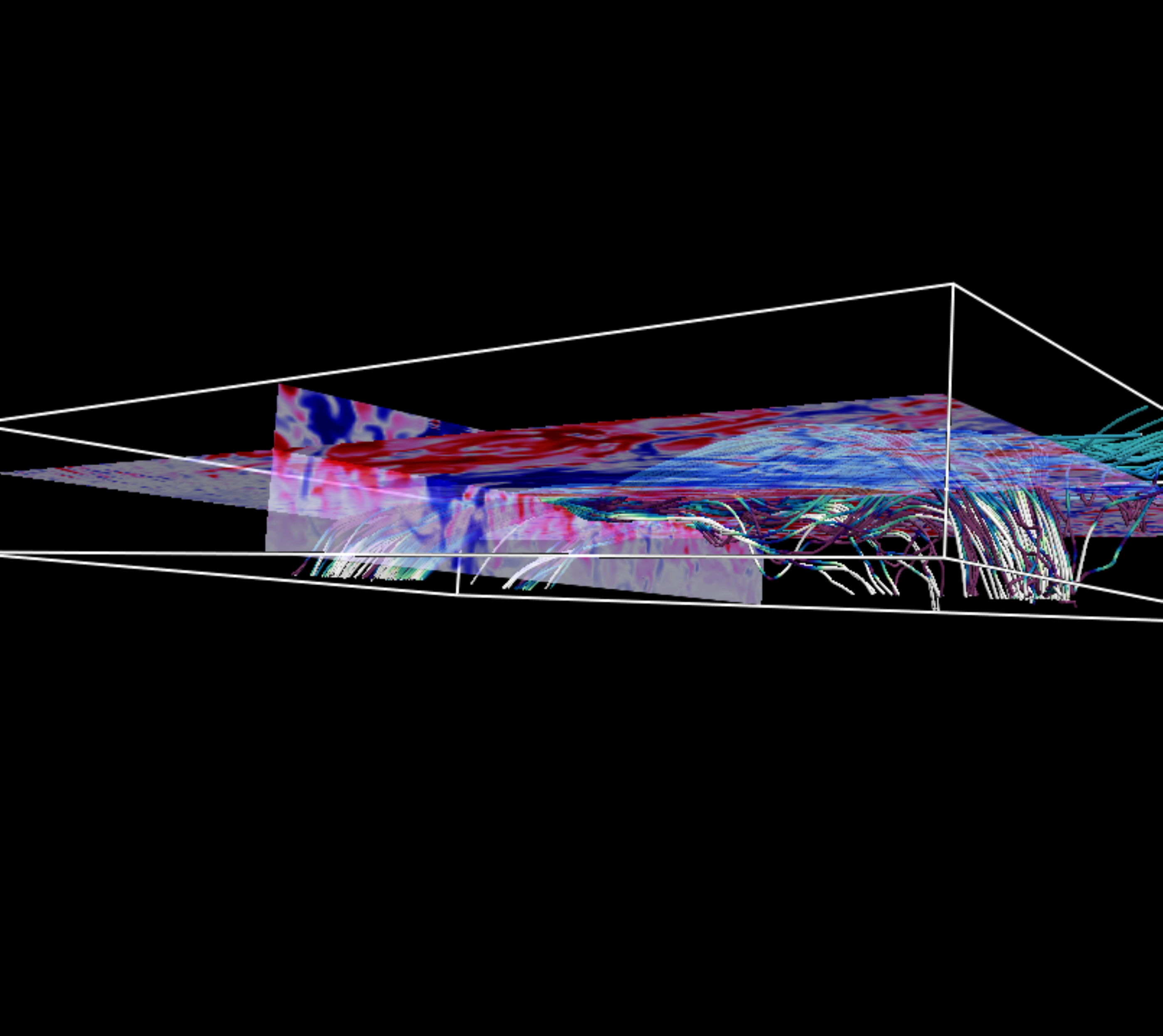} 
    \caption{Field-line connectivity and associated velocity and pressure perturbations in the simulated penumbra: CEF filaments (left figures) and NEF filaments (right figures). In all figures the semi-transparent horizontal and vertical planes show
the field-aligned velocity component from $\pm 10$ km s$^{-1}$ near the average $\tau=1$ level, outflows are red and inflows are blue. The vertical
plane is placed close to the inner penumbral boundary and is used to select field lines by their flow velocity.
In the upper figures, the color of the field lines shows the flow velocity along
the field. In the bottom figures, the color of the field lines indicates
 the pressure perturbation along the field lines from $\pm 10^5$ dyne/cm$^2$, where white is low and purple is high pressure. The upper and lower panels refer to different view angles. The plots correspond to the snapshot at $t=65$ hr.}\label{fig:11}
\end{figure*}

Unlike the NEF, the CEF is driven at all times by the contribution of two different radial forces: $P_r^-$ and $L_r^-$.
Figure \ref{fig:10} shows the temporal evolution of the average radial mass flux in the near-surface layers (from $z=-500$ to 0 km) at the sources of the CEF in the selected penumbral portion (left plots) and, for comparison, at the sources of the NEF (right plots).
The plotted time interval covers the  lifetime of the CEF (i.e. the time from 60 to 70 hr). The solid lines  in Figure \ref{fig:10} show the average radial energy conversion terms in the near-surface layers at each time-step.
 The temporal evolution of the CEF shows that the negative radial mass flux (dashed line) is strongly variable compared to the NEF case, and that it is proportional to the radial acceleration, $A_r^-$.
This means that  the sum of the two radial forces, magnetic and pressure forces (note that $A_r^- \approx -(L_r^-+P_r^-$)), is responsible for modulating the inward mass flux of the CEF in the near surface layers.

Unlike the CEF, the positive radial mass flux  of the NEF as well as the radial energy balance have a more steady behavior during the 10 hours of the analyzed interval, exhibiting the quasi-stationarity and robustness of the NEF feature when the penumbra is well-developed. The radial Lorentz terms 
keep in balance with the radial acceleration forces during the 10 hours here analyzed and are, consistently, the responsible forces driving the NEF in the radial direction.

\section{Field-line Connectivity}

We have shown in the previous analysis that the CEF and the NEF are both driven within a thin boundary layer close to the $\tau=1$ level, and that the velocity vector is mostly radial. Moreover, we found that strong radial pressure gradients exist everywhere within the thin boundary layer in which the CEFs are driven. In contrast, while the radial pressure forces have an on average opposed contribution to the driving of the NEF in the source regions, positive  radial pressure forces dominate the average in the outer penumbra where the NEF sinks (Figure \ref{fig:8}).
Although the driving of the flows depends principally on the conditions in their respective upflow cells, identifying the magnetic connectivity of the field lines is an important aspect to describe individual flow channels (penumbral filaments) as flux tubes, and therefore is also necessary to validate whether the flows are or are not driven by a siphon flow mechanism. Nonetheless, such a representation involves some limitations when compared with models based on the thin flux tube approximation \citep[e.g.,][]{Meyer1968,Thomas1993, Montesinos1997,Schlichenmaier1998a} which assume well-defined footpoints of the flux tubes on horizontal cuts made at given heights below the surface. In particular,  the NEF-carrying filaments in this simulation  \citep[and in that by ][]{Rempel2011b} display elongated upflow cells which constrain their representation under the thin flux tube picture.

To study the magnetic field-line connectivity as well as the associated field-aligned velocity component and pressure perturbation along the different types of filaments in the penumbra (NEF carrying and CEF carrying filaments), we use the
VAPOR software package developed at NCAR \citep[][http://www.vapor.ucar.edu]{Clyne2005,Clyne2007}. 

In Figure \ref{fig:11}, we have selected field lines in a cross-section 
perpendicular to filaments based on regions with strong horizontal flows. The selected  field lines were colored according to their field-aligned velocity in the upper images and  according to the gas pressure perturbation along the field lines in the bottom images (we compute the pressure perturbation after subtracting the hydrostatic
mean, since stratification dominates).
The images on the left show field lines carrying a CEF from two different view angles and the images on the right show field lines  carrying a NEF. In the case of the CEF filaments, most of the field lines span 
from the outer to the inner penumbra with a consistent inflow
along them (top left panel) and they have a very systematic pressure
perturbation (bottom left panel) with the gas flowing from high pressure regions (purple) to low pressure regions (white), consistent with a siphon flow.
 
In the case of the NEF, the velocity along the field lines is less
consistent and some field lines have even an inverse flow in their upper portion
 (top right panel), similarly to the findings of \citet{Rempel2011b}. In particular, in the inner penumbra, field lines
that host the fastest Evershed flows
still connect to the top boundary, but forming small dips in-between their extremes. 
This
indicates that the NEF is to a lesser degree a flow along the field and it implies
continuous reconnection and change of field line connectivity as the
mass moves outward in the penumbra. 
Moreover, it implies that the physical conditions in the outer footpoint have a minor impact on the driving of the NEF.
Looking at the pressure perturbation 
(bottom right), we find a mix of high and low pressure at both
footpoints. There could still be siphon flows along some field lines (mostly in the outer
penumbra), but they do not provide a consistent explanation for the
Evershed flow in the bulk of the penumbra.

\section{Discussion and conclusion}

We presented a detailed analysis of the properties of penumbral fine structure associated with the counter Evershed flows (CEF) and the normal Evershed outflows (NEF) near $\tau=1$ in the penumbra of the recent numerical sunspot simulation by \citet{Rempel2015}. Our investigation mainly focuses on the physical driving mechanisms of the NEF and the CEF in the near surface layers of this simulation.

The main difference found between the penumbral filaments carrying the NEF and the ones carrying the CEF is the location within the penumbra and the radial extent of the upflow cells. Both aspects are crucial for the driving of each type of flow given that their driving mechanisms depend primarily on the conditions present in their respective upflow cells. 

The energy conversion analysis at the sources of both, outflows and inflows, suggests that the CEF and the NEF are both driven within a thin boundary layer close to the $\tau=1$ level, and no substantial driving forces exist well above $\tau=1$. In both cases, the loss of buoyancy due to radiative cooling is seen as a change of sign of the work by pressure in the vertical direction close to $z=0$ (Figure \ref{fig:6}). However, while this provides only a small contribution to the deceleration of the vertical flow in the case of the NEF, i.e. most of the deceleration happens deeper down where the pressure driving is still positive and the dominant offset is given by the vertical Lorentz force, the deceleration of upflows by loss of buoyancy seems to be more prominent for the CEF near $z=0$. Nonetheless,  deceleration of upflows in CEFs near $\tau=1$ occurs mainly due to an opposed  contribution from the vertical Lorentz force, similarly to the NEF case. Therefore,  notable differences exist with respect to the underlying driving forces of both, NEF and CEF, compared to field-free convection where deceleration of vertical flows near the surface is mainly due to loss of buoyancy caused by the radiative cooling \citep[see][who analyzed in detail the driving of the NEF in a simulated sunspot penumbra and made a comparison between the NEF and the plage region surrounding the sunspot, and found that the vertical pressure/buoyancy breaking has  very clear distinct signatures in granulation and in the penumbra]{Rempel2011b}.

For the driving of the NEF, we found similar results to those reported by \citet{Rempel2011b}. 
There is an almost complete balance between vertical pressure and Lorentz forces in upflows, as well as between the Lorentz force and acceleration  in the radial direction.
We found no significant kinetic energy in the vertical flow component at the sources of the NEF but a strong acceleration in the radial component in the near-surface layers. This implies that the flow changes direction and gains at the same time significant amounts of kinetic energy in the radial direction. 
Since the Lorentz force does not do any net work but only changes the direction of the flow, we can say that the flow is deflected by the highly inclined magnetic field in the penumbra and the gas is accelerated in the radial direction as a result of mass flux conservation coupled with the low gas density at the surface.
The most important feature driving outflows in the inner and middle penumbra is the  increase of the radial field component close to $\tau=1$.
These results strongly support the magneto-convective driver scenario of the NEF, as proposed in 
\citet{Scharmer2008}, where the NEF is described as the horizontal component of overturning convection in penumbrae.

It has also been proposed in some models that the NEF could be the result of stationary siphon flows driven by a gas pressure difference between the footpoints of the penumbral filaments \citep[e.g.][]{Meyer1968}. Such models consider processes related to turbulent pumping near the outer penumbral boundary \citep{Montesinos1997,Brummell2008} which make the field lines  bend downwards, and additionally produce a field strength enhancement in the outer filament endpoint due to strong downdrafts \citep{vannoort2013}. Such a field enhancement would establish a gas pressure gradient between the filament endpoints, producing a siphon-like outflow along the field lines.
We do observe a net field strengthening at the outer endpoints of the NEF-carrying filaments (see e.g. Figure \ref{fig:4}), at the places where the NEF sinks in the outer penumbra. This enhanced field could explain the increase of the radial pressure gradient towards the outer penumbra seen in the bottom right plot in Figure \ref{fig:8}.
However, the radial pressure gradients do not play an important role in the inner and in the middle penumbra, where most of the driving of the NEF occurs. This can be attributed to the elongated aspect of the upflow cells, which produces a substantial reduction of the radial pressure gradient.
The fact that the largest average outflow speeds  in the penumbra ($>5$ km s$^{-1}$) stand out towards the outer penumbra (see radial velocity profile of the outflows in Figure \ref{fig:8}) might be in part due to a larger fill factor of outflows in the outer penumbra that dominates the average.
According to our results, the overall picture is that of a NEF mainly driven by magneto-convection in the inner and middle penumbra, in agreement with the results of \citet{ Rempel2011b}. 

Observational studies of sunspots have shown that the penumbral filaments carrying the NEF in penumbrae have well defined "footpoints" at $\tau=1$, see e.g. \citet{Tiwari2013}. The filaments footpoints are seen to confine the upflows (sources) and the downflows (sinks) associated to the Evershed flow in the penumbra, separately. 
This is an important difference given that in these simulations the upflow cell covers a large portion of the NEF filaments, meaning that almost the entire filaments behave as a "footpoint".
 However, the lack of information on a true geometrical height scale in the current observational analysis techniques makes the comparison of numerical models with the highest-resolution observations very challenging, unless the simulations include the computation of the observational quantities using radiative transfer computations (forward computations). 

Overall, our analysis of the NEF essentially reinforces conclusions of \citet{Scharmer2008},  \citet{Rempel2009a} and \citet{ Rempel2011b} stating that the penumbra is dominated by
anisotropic magnetoconvection and that the NEF can
be understood as the convective flow component in the direction
of the magnetic field, i.e., the overall underlying energy source is convective instability.

Unlike the NEF, the driving of the CEF 
occurs mainly in the middle and outer penumbra 
and the footpoints (sources and sinks) of the CEF-carrying filaments are generally well separated from each other. 
Furthermore,  the upflow cells in the CEF-carrying filaments are less spread out compared to the NEF, i.e. less elongated in the radial direction, which
enhances the role of the radial pressure gradients  and allows for the existence of
siphon-like inflows along the field lines that are connecting to the inner penumbra.

In addition to the radial pressure gradients, the CEF 
is also affected by the inclined field of the penumbra, in a similar way as the NEF is, i.e. that the radial Lorentz force plays an important role in the driving of the CEFs.
The combination of both radial forces,  results in fast inflows ($>5$ km s$^{-1}$ on average) near the $\tau=1$ level.
We found that, irrespective of whether the sources of the CEF are located in the middle or in the outer penumbra, the upflowing gas suffers a  reduction of the vertical magnetic field 
combined with an enhancement of the radial field component. This leads to a positive contribution of the radial Lorentz force for the driving of the CEF. 

Nonetheless, the energy conversion associated to the radial Lorentz force is on average smaller in the CEF sources than in the NEF sources given that the magnetic field is on average weaker in the outer penumbra than in the inner penumbra. This would at first sight imply a smaller radial acceleration of the CEF compared to the NEF. However, it is generally the sum of both, the radial pressure force and the radial Lorentz force that determines the radial acceleration of the CEF. Consequently, the inward acceleration of the fluid can eventually become even larger than that of the NEF. This occurs at several stages of the evolution of the CEF in the analyzed simulation.
However, during the 10 hours analyzed in these simulations, the CEF shows up as a highly variable and unstable flow, while the NEF appears as a quasi-stationary and more robust feature in the penumbra.

Both, the CEF and the NEF are strongly magnetized and are driven within a thin boundary layer close to $\tau=1$. There are also other, smaller scale flows (outflows and inflows) in the penumbra that are unrelated to the filamentary penumbra. In particular, the additional inflows are mostly driven by pressure gradients in regions with an (on average) opposed contribution of the radial Lorentz term, the latter favoring outflows. This could explain why these inflows are of much smaller scales than the CEF.

According to  observational studies, CEFs can occur under different physical circumstances. 
On the one hand, those CEFs that have been observed during the early stages of penumbra formation \citep[e.g.,][]{ Schlichenmaier2012,Romano2014,Murabito2016} usually appear as elongated patches at the outer edge of the proto-spot, and are unrelated to any filamentary structure since they are observed when the latter 
is not yet developed.
Furthermore,  
the newly formed  penumbral  filaments host a NEF soon after their formation. Therefore, those inflows observed around forming sunspots may involve different physical driving mechanisms since they are essentially different to the CEFs studied in this simulation. A possible scenario is that they
are  driven only by gas pressure gradients  which are caused by the increase of the magnetic field strength in the proto-spot.

On the other hand, the CEFs that have  been observed (albeit more rarely) in well-developed penumbrae  \citep{Kleint2013,Louis2014,Siu2017} share important similarities with the CEFs in the present simulations: they were observed either along singular penumbral filaments or, as in \citet{Siu2017}, along an array of penumbral filaments covering a sizable part of
 the penumbra in a mature spot, but with most of the penumbra still displaying the NEF as in our present simulation. This occurs in all the three above-mentioned reported observations of CEFs in well-developed penumbrae.
However, due to the lack of knowledge of
the true geometrical height scale in the observations and the inability to measure vertical gradients below the photosphere, none of these works could determine the dominant forces driving the flows.

In particular,
\citet{Siu2017} reported the observation of a prominent CEF with lifetimes of $\sim 2$ days at photospheric heights. Similar to the simulations analyzed here, the CEF in those observations showed an associated filamentary structure, with the sources  of the CEF (hot upflows in vertical field regions) identified in the outer penumbra and the sinks (cooler downflows in strong vertical field regions) at the inner penumbral boundary. 
Furthermore, we found that the general magnetic, thermal and velocity structure along the central axes of the CEF-carrying filaments at $\tau=1$ are remarkably similar  to those reported in that work. Of particular interest is their finding of an enhanced temperature  at the sinks of the CEF with respect to the surrounding environment and a very large field strengthening associated to supersonic downflow speeds. We also see these two aspects at the sinks of the CEF in the simulations. On the one hand, we found that deceleration takes place in the form of shocks at the sinks of both, CEF- and NEF-carrying filaments. This can explain the local temperature rise found at their sinks.
On the other hand, we found enhanced field strengths at the sinks of both, CEF and NEF, reaching values up to $\sim 5$ kG and $\sim2.5$ kG respectively. The very low densities of the downflowing gas produces the depression of the local $\tau=1$ level at the sinks of both, NEF and CEF. At the CEF sinks, the $\tau=1$ depression combined with the influence of the umbral field contributes to the observation of such strong fields at those places; while at the NEF sinks, the local enhancement of the field is the result of the $\tau=1$ depression combined with a net magnetic field intensification that is well-localized in height and might be produced by the supersonic downdrafts of magnetic flux at the outer penumbra, as proposed by \citet{vannoort2013}.

The CEF in these simulations persists up to $\sim10$ hours, which is much shorter than the lifetimes reported by \citet{Siu2017}, although  the feature studied by \citet{Siu2017} was extraordinary compared with other observed CEF events \citep[see e.g.][]{Kleint2013,Louis2014}. However, in both cases, the CEF has much shorter lifetimes than the NEF and can be thought of as transitory events relative to the latter. Furthermore, similar to what happens in these simulations, \citet{Siu2017} reported that the penumbral sector harboring the CEF displays only a NEF  after the CEF disappears.  We have not studied the exact mechanism  how the CEF is reversed to a NEF. 
However, a negative contribution of the radial Lorentz term combined with a strong reduction of the radial pressure gradient 
at the sources of the CEF would strongly favor the driving of outflows instead. A physical mechanism leading to the change of the magnetic field configuration and of the plasma conditions in the penumbra in such a way would likely involve magnetic reconnection.

Our analysis of the driving forces of the CEF in the simulated penumbra clearly shows that the CEF is a siphon flow driven by pressure gradients along the penumbral filaments. The inclined field in the penumbral filaments causes the radial Lorentz force to also play an important role in accelerating the gas inwards. Furthermore, besides the most prominent and persistent groups of CEF-carrying filaments during the analyzed temporal interval (e.g. those shown in Figure \ref{fig:2} and the portion analyzed in Figures \ref{fig:9} and \ref{fig:10}), there are also a few other CEF-carrying filaments that live shorter and appear intermittently at different azimuths  in the penumbra. 
We do not discard the possibility that CEFs may actually occur along penumbral filaments more regularly than observed, but their short lifetimes prevents them being easily observed. According to our results, that would occur whenever a significant pressure gradient favoring CEFs is established in a penumbral filament, as in those cases in which the outer footpoint of the filament is not strengthened with respect to the inner footpoint (see, e.g., the CEF-carrying filament shown in Figure \ref{fig:4}). Alternatively, CEFs may also appear as a consequence of ongoing magnetic flux emergence in the penumbra.  This interpretation was presented by \citet{Chen2017}. In their recent MHD simulation more than 60$\%$ of the simulated penumbra is dominated by CEFs that result from the mass drain into the umbra-penumbra boundary along the newly emerged field lines that coalesce and contribute to the horizontal field in the penumbra.

Nonetheless, there are very few reported observations of CEFs in sunspot well-developed penumbrae \citep{Kleint2013,Louis2014,Siu2017}. This could also be due to a combination of their rare  occurrence and their short lifetimes.
However, more theoretical studies on this topic as well as 
high-temporal-cadence spectropolarimetric observations are necessary to investigate the true occurrence frequency and to learn more about the nature of CEFs in well developed sunspot penumbrae as well as their possible influence on the upper atmosphere.

The inverse Evershed flow (IEF) in the uppermost part of the simulation box is a robust feature and the question of an IEF-CEF connection is compelling. However, the IEF shown in the simulation is possibly affected by the proximity of that flow to the top boundary and by 
 Alfv\'en speed
reduction. This will be addressed
when improved simulations become available.

\acknowledgments
Acknowledgments.

This work was carried out in the frame of the International
Max Planck Research School (IMPRS) for Solar System Science at the Max
Planck Institute for Solar System Research (MPS). It is supported by the Max
Planck Society and by BECAS CONACyT AL EXTRANJERO 2014.
This work was partly supported by the ERC under grant agreement No. 695075 and the BK21 plus program through the National Research Foundation (NRF) funded by the Ministry of Education of Korea. The National Center for Atmospheric Research is sponsored by the National Science Foundation.

\bibliographystyle{yahapj}
\bibliography{references}


\end{document}